\documentclass[sigconf]{acmart}

\usepackage{colortbl}      
\usepackage[table]{xcolor}
\usepackage{amsmath}
\usepackage{latexsym}
\usepackage{multirow}
\usepackage{enumitem}
\usepackage{makecell}
\usepackage{mdframed} 
\usepackage{ragged2e} 
\usepackage{algorithm}
\usepackage{algpseudocode}
\usepackage{arydshln}
\usepackage{enumitem}
\usepackage{balance}

\usepackage{amsthm}
\usepackage[english]{babel}

\newcommand{\miniskip}{\vspace*{-.5\baselineskip}}
\newcommand{\shrink}{\vspace*{-.9\baselineskip}}

\newcommand{\rrrbaseline}{RrrC}
\newcommand{\reabaseline}{ReaC}
\newcommand{\selfbaseline}{SelfC}
\newcommand{\proposedue}{$\text{R}^2$C} 
\newcommand{\MainTask}{Uncertainty Quantification}
\newcommand{\MainTaskAbbr}{UQ}

\AtBeginDocument{%
  }

\copyrightyear{2026}
\acmYear{2026}
\setcopyright{cc}
\setcctype{by}
\acmConference[SIGIR '26]{Proceedings of the 49th International ACM SIGIR Conference on Research and Development in Information Retrieval}{July 20--24, 2026}{Melbourne, VIC, Australia}
\acmBooktitle{Proceedings of the 49th International ACM SIGIR Conference on Research and Development in Information Retrieval (SIGIR '26), July 20--24, 2026, Melbourne, VIC, Australia}
\acmDOI{10.1145/3805712.3809717}
\acmISBN{979-8-4007-2599-9/2026/07}

\begin{document}

\title{\MainTask\ for Retrieval-Augmented Reasoning}

\author{Heydar Soudani}
\affiliation{%
  \institution{Radboud University}
  \city{Nijmegen}
  \country{The Netherlands}}
\email{heydar.soudani@ru.nl}

\author{Hamed Zamani}
\affiliation{%
  \institution{University of Massachusetts Amherst}
  \city{Amherst}
  \country{United States}}
\email{zamani@cs.umass.edu}

\author{Faegheh Hasibi}
\affiliation{%
 \institution{Radboud University}
 \city{Nijmegen}
 \country{The Netherlands}}
\email{faegheh.hasibi@ru.nl}

\renewcommand{\shortauthors}{Heydar Soudani, Hamed Zamani, and Faegheh Hasibi}

\begin{abstract}

Retrieval-augmented reasoning (RAR) is a recent evolution of re\-trieval-augmented generation (RAG) that employs multiple reasoning steps for retrieval and generation. While effective for some complex queries, RAR remains vulnerable to errors and misleading outputs. Uncertainty quantification (UQ) offers methods to estimate the confidence of systems' outputs. These methods, however, often handle simple queries with no retrieval or single-step retrieval, without properly handling RAR setup. Accurate estimation of UQ for RAR requires accounting for all sources of uncertainty, including those arising from retrieval and generation. In this paper, we account for these sources and introduce Retrieval-Augmented Reasoning Consistency (\proposedue), a novel \MainTaskAbbr\ method for RAR. The core idea of \proposedue\ is to perturb the multi-step reasoning process by applying various actions to reasoning steps. These perturbations alter the retriever’s input, which shifts its output and consequently modifies the generator’s input at the next step. Through this iterative feedback loop, the retriever and generator continuously reshape each other’s inputs, enabling us to capture uncertainty arising from both components. Experiments on five popular RAR systems across diverse QA datasets show that \proposedue\ improves AUROC by over 5\% on average compared to the state-of-the-art \MainTaskAbbr\ baselines. Extrinsic evaluations using \proposedue\ as an external signal further confirm its effectiveness for two downstream tasks: in the Abstention task, it achieves \textasciitilde5\% gains in both F1Abstain and AccAbstain; in Model Selection, it improves exact match by \textasciitilde7\% over single models and \textasciitilde3\% over selection methods. 
Code is available on \href{https://github.com/HeydarSoudani/R2C}{https://github.com/HeydarSoudani/R2C}.

\end{abstract}




\maketitle

\section{Introduction}~\label{sec:intro}

\begin{figure*}[t]
  \centering
  \includegraphics[width=0.94\textwidth]{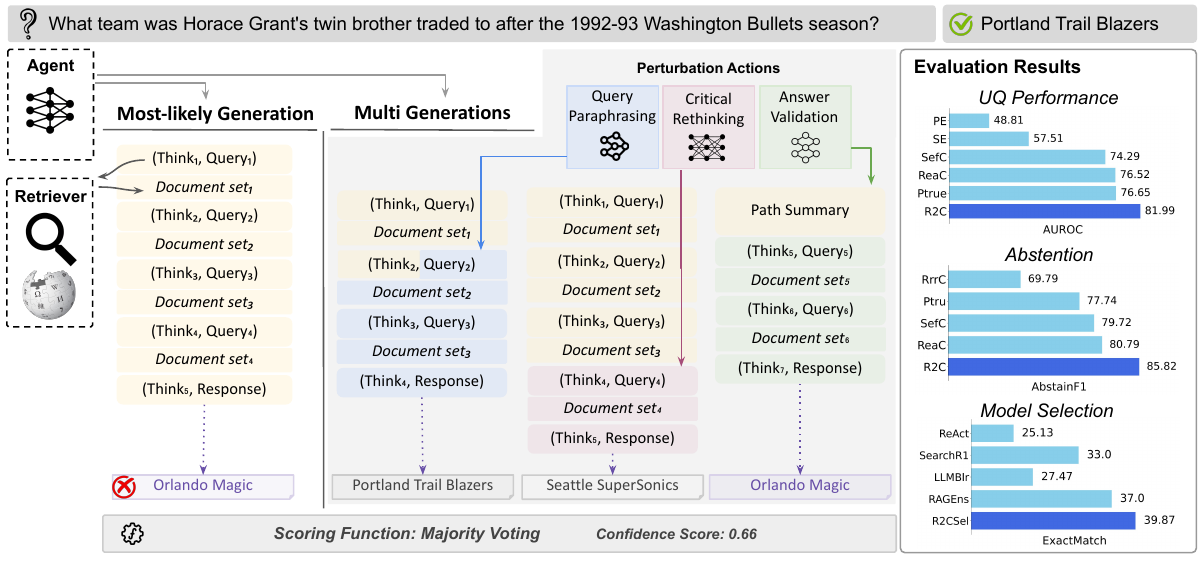}
  \shrink
  \caption{
  \proposedue\ overview. Given a user query, the agent (LLM) first generates the most-likely reasoning path leading to the most-likely response (left, yellow). To estimate uncertainty, \proposedue\ creates multiple perturbed generations by randomly altering states in the reasoning path (middle, gray). The confidence score is then derived via majority voting.  \proposedue\ significantly outperforms established \MainTaskAbbr\ methods and achieves significant improvements on two downstream tasks: abstention and model selection.
  }
  \label{fig:main_fig}
   \shrink
\end{figure*}

Retrieval-augmented generation (RAG) is widely used for know\-ledge-intensive tasks, but remains limited in addressing complex multi-step reasoning~\cite{Jin25SearchR1, Hoveyda:2025:AOM, Survey24Soudani, Salemi24Evaluating, Hoveyda:2026:orlog}. Recent work has explored combining RAG with reasoning, where LLMs are prompted or trained to use search engines as tools during their reasoning process; a paradigm referred to as retrieval-augmented reasoning (RAR)~\cite{Sun25ReARTeR, Tran25RARE, Islam24OpenRAG, Jin25SearchR1}. However, RAR models are still prone to producing incorrect responses, due to issues such as retrieving irrelevant documents in early steps, misinterpreting retrieved content, or misusing internal knowledge. Therefore, ensuring the trustworthiness of RAR outputs has become a critical challenge.

Uncertainty quantification (UQ) is a widely studied task in machine learning, aimed at assessing the reliability of model outputs by measuring the degree of uncertainty (or lack of confidence)\footnote{
Some literature distinguishes between a model’s uncertainty and its confidence in a response~\cite{lin24bbox}. In this work, we follow the terminology used in most of the literature, where uncertainty is treated as a lack of confidence and the two terms are used interchangeably~\cite{liu2025uncertainty}.} 
a model has in its predictions~\cite{Malinin21LNPE, Kadavath22PE, Kuhn23SE, Hou24Decomposing, Ye24Benchmarking, Harrison24Variational}. 
Recent methods of estimating the uncertainty of LLM outputs are designed for settings where the input consists solely of a query, meaning the LLM itself is the only source of uncertainty~\cite{Bakman24mars, yaldiz2025lars, duan2024sar}. The limited work on UQ for RAG~\cite{soudani2025uncertainty, perez2025uncertainty} incorporates the document–response relationship into the uncertainty score, but these methods are only applicable for simple RAG settings, where documents are retrieved once and inserted into the input prompt for generation. As a result, existing UQ approaches are suboptimal for RAR.

A fundamental reason for the relatively poor performance of existing UQ methods for RAR models is that they primarily attribute uncertainty to the LLM's generative process; i.e., next-token prediction. In RAR systems, however, 
we have more sources of uncertainty:
the \textit{retriever}, which may provide irrelevant or partially relevant retrieved documents and potentially mislead the model's reasoning and response-generation processes; and the \textit{generator}, where the model's reasoning may deviate from the user query's intent and  retrieved documents, leading it to formulate new search queries that fail to gather informative evidence.

In this paper, we propose \textbf{R}etrieval-Augmented \textbf{R}easoning \textbf{C}on\-sist\-ency (\proposedue), a novel \MainTaskAbbr\ method that, unlike previous approaches, accounts for multiple sources of uncertainty in RAR. 
The central idea of \proposedue\ is to allow the model to explore diverse reasoning paths, queries, and documents and then measure the consistency of the resulting final answers. This is achieved by modeling RAR as a Markov Decision Process (MDP) and perturbing this process in a controlled way  through a set of perturbation actions across various states. Three perturbation actions are designed to influence query generation, document retrieval, and LLMs' thinking process. 
These perturbations enable models to arrive at diverse final responses for uncertain generations. 
The confidence score is then obtained by measuring consistency of the generated answers using majority voting; see Figure~\ref{fig:main_fig}.

We conduct our experiments across multiple datasets and five RAR models. Our experiments show  that \proposedue\ significantly outperforms existing LLM-specific \MainTaskAbbr\ methods, achieving on average more than a 5\% improvement in AUROC compared to the state-of-the-art UQ methods.
We further extrinsically evaluate \proposedue\ on two downstream tasks: 
(i) Abstention:\footnote{Also referred to as selective prediction in the literature~\cite{Duan25UProp, Xin20Selective}.} the task of generating `I don't know' when the model is uncertain about its output~\cite{Feng24Hallucinate, MadhusudhanMYH25},
and (ii) Model Selection:\footnote{Also referred to as selection-based model ensemble in the literature~\cite{Jiang23LLMBlender}.} the task of selecting a final answer from a pool of candidates generated by multiple systems~\cite{Chen25Harnessing, GuhaCCKR24}.
Our experimental results indicate that \proposedue\ delivers statistically significant gains over existing approaches: 
in Abstention, it achieves roughly 5\% improvements on both F1Abstain and AccAbstain;
in Model Selection, it increases exact match by about 7\% relative to single RAR models and by about 3\% compared to selection model baselines.

Given the strong performance of \proposedue\ in both direct evaluation and extrinsic evaluation on downstream tasks, we investigate the factors that contribute to its effectiveness. 
We show that \proposedue\ retrieves on average 25 unique documents for each score, compared to 16 documents retrieved by other \MainTaskAbbr\ methods. It also achieves a query diversity of $0.35$ compared to $0.30$ of other methods, measured by the inverse of the average pairwise cosine similarity between queries~\cite{cox2021directed, zhang2025evaluating}.
This diversity in queries and documents demonstrates that \proposedue\ generates diverse, yet relevant reasoning paths through our controlled perturbation mechanism.  
As a result, this enables the method to achieve confidence scores comparable to baseline approaches while requiring only about 3 generations on average, 2.5 times fewer token generations than the 10 used by the baselines.
This highlights that \proposedue\ is not only the most effective method of its kind but also a relatively more efficient \MainTaskAbbr\ approach.
%
To summarize, the main contributions of this paper are: 
\begin{enumerate}[leftmargin=*]
    \item We propose \proposedue, a novel theoretically grounded UQ method based on MDP; the first of its kind that captures different sources of uncertainty in RAR.
    \item We conduct extensive experiments on three datasets and five RAR methods, demonstrating the superiority of the proposed method on the UQ task with average AUROC of 82\%.
    \item We show the effectiveness of our method on both model selection and abstention tasks, significantly outperforming baselines by at least 3\%.
    \item We demonstrate that \proposedue\ achieves an improvement in token efficiency by approximately 2.5 times.
    \item We show that diverse query and document generation strengthens UQ by capturing multiple uncertainty sources.

\end{enumerate}

\section{Related Work}~\label{sec:related_work}

\noindent
\textbf{Retrieval-Augmented (Reasoning) Models.}
RAG is a framework that combines the strengths of retrieval models and generative models~\cite{Noise2024Florin, Soudani24FTvsRAG}. Broadly, RAG can be implemented in different ways. In the retrieve-then-generate paradigm, relevant documents are first retrieved based on the user's input and then incorporated into the model's prompt~\cite{mallen23popqa, Soudani24FTvsRAG}. In contrast, Active RAG allows retrieval to occur throughout the generation process, either in fixed intervals or dynamically, whenever additional information is needed~\cite{trivedi2023ircot, jiang2023flare, su2024dragin}.
Retrieval-Augmented Reasoning (RAR) is a recent extension of RAG that integrates retrieval with reasoning, aiming to improve the interaction between LLMs and retrievers~\cite{Sun25ReARTeR, Tran25RARE}. 
For example, SelfAsk~\cite{press2023selfask} decomposes complex questions into follow-up queries and intermediate answers, while ReAct~\cite{Yao23ReAct} defines a set of actions, such as search, lookup, and finish, to structure interactions with external resources. 
More recent and effective models such as ReSearch~\cite{chen25research} and Search-R1~\cite{Jin25SearchR1} are explicitly trained to seamlessly integrate external resources into reasoning.
Despite their effectiveness on complex queries, RAR models remain prone to errors, such as retrieving irrelevant documents, misinterpreting content, or misusing internal knowledge.

\noindent
\textbf{\MainTask\ (\MainTaskAbbr) and Confidence Estimation (CE).}
\MainTaskAbbr\ and CE are closely related but conceptually distinct. In traditional machine learning, uncertainty is defined as a property of the model’s predictive distribution  given a specific input~\cite{lin24bbox, liu2025uncertainty}. 
Confidence, by contrast, reflects the model’s belief in the correctness of a \textit{particular prediction} for the given input.
In other words, CE is defined as the task of quantifying how certain a model is about a specific generated response~\cite{liu2025uncertainty}, while UQ is the task of capturing the degree of variability or unpredictability in the model’s outputs (irrespective of a specific response)~\cite{kendall2017uncertainties, gawlikowski2023survey}. Despite this distinction, existing work sometimes uses uncertainty to refer to estimated confidence~\cite{Kadavath22PE, liu2025uncertainty}. In this paper, we compute confidence scores for LLM-generated responses. Nevertheless, following common practice in the literature~\cite{Kadavath22PE, Malinin21LNPE, wang2024conu, ren2022out, fadeeva2024fact, yin2024reasoning, zhang2025cot, becker2024cycles, liu2024uncertainty}, we refer to this confidence score as an uncertainty measure and compare it against other \MainTaskAbbr\ methods.

Broadly, existing UQ methods can be divided into two categories: white-box approaches, which leverage token-level probabilities and entropy~\cite{Kadavath22PE, Kuhn23SE, Bakman24mars, yaldiz2025lars, duan2024sar, Soudani25Enhancing}, and black-box approaches, which rely only on the final textual outputs~\cite{lin24bbox, Tian23Verbalized}.
Most \MainTaskAbbr\ methods focus on question answering and view the LLM as the only source of uncertainty, but in RAG, the retriever also contributes its own uncertainty, making this assumption incomplete.
Limited research has explored \MainTaskAbbr\ for RAG by modeling the document-response link, either via axioms~\cite{soudani2025uncertainty} or utility models~\cite{perez2025uncertainty}. 
Some path-based approaches focus on assessing the consistency of reasoning paths for reasoning tasks~\cite{Wang23SelfConsistency, qi2025rstar, li2025raspberry}.
However, these methods do not extend naturally to RAR that involves repeated retrieval during reasoning.
Recent work has studied uncertainty propagation in multi-step decision-making by combining uncertainties from intermediate steps. SAUP~\cite{Zhao25Propagation} learns aggregation weights to merge per-step uncertainties, but it relies on ground-truth labels from the test domain.
In contrast, we propose \proposedue, a method that accumulates uncertainty over the entire reasoning path, while considering different sources of uncertainty including the retriever and the generator.
%

\section{Preliminaries}~\label{sec:preliminary}

\setlength{\abovedisplayskip}{2pt} 
\setlength{\belowdisplayskip}{2pt} 

\begin{algorithm}[t]
\caption{\textbf{\proposedue}: Retrieval-Augmented Reasoning Consistency}
\label{alg:r2c}
\begin{algorithmic}[1]
\Require user query $x$, backbone LLM $\pi_\theta$, 
number of generations $B$, set of main actions $A = \{a_{ret},a_{ans}\}$, set of perturbation actions $A^* = \{a_{qp},a_{cr},a_{av}\}$
\Ensure Confidence score $C(x, r)$ 
\State \textbf{for} $t = 1$ \textbf{to} $N$ \textbf{do} $s_t \gets \pi_{\theta}\!\left(s_{t-1}, a_{t-1}\right)$ \Comment{Generate the most-likely path iteratively}

\State $r \leftarrow s_N$ \Comment{Capture the most-likely response} 

\State $R_G = \emptyset$  \Comment{Initialize multi-generation response set}
\For{$b = 1 \to B$}
    \State $a^* \sim \mathcal{U}(\mathcal{A}^*)$  \Comment{Sample an action from $\mathcal{A}^*$}
    \If{$a^* = a_{av}$} {$s_t \gets s_N$}   \Comment{Select the last state for $a_{av}$}
    \Else { $s_t \sim \mathcal{U}(s_1, s_{N-1})$ } \Comment{Sample a state}
    \EndIf 
    
    \State $s_{t+1} = \pi_{\theta}(s_t, a^*)$ \Comment{Apply action $a^*$ at state $s_t$}
    \State \textbf{for} $i = 1$ \textbf{to} $N^b$ \textbf{do} $s_{t+i+1} \gets \pi_{\theta}\!\left(s_{t+i}, a_{t+i}\right)$ 
    
    \State $r^b \leftarrow s_{N^b}$ \Comment{Capture the sampled response}
    \State $R_G = R_G \cup \{r^b\}$ \Comment{Update the sampled response set}
\EndFor
\State $c = \;\frac{1}{B}\sum_{b=1}^{B}\mathbb{I}(r^b = r)$ \Comment{Compute the confidence score}
\State \Return $c$ 
\end{algorithmic}
\end{algorithm}
 \shrink

\noindent
\textbf{RAR as Markov Decision Process.} 
We formalize RAR as a stochastic Markov Decision Process (MDP), described by a quadruple $(S, A, P, R)$, where $S$ denotes a set of states, $A$ represents a set of actions the agent can take in a state, $P(s_{t+1} \mid s_t, a_t)$ denotes the probability of transitioning from state $s_t$ to state $s_{t+1}$ given action $a_t$, and $R( s_t, a_t)$ is the reward received by the agent after taking action $a_t$ in state $s_t$.
%
To generate a factual response to a user query $x$, the agent $\pi$ starts from an initial state $s_0$ corresponding to the query $x$. The agent then iteratively selects an action at each step until it chooses a halting action and generates the response. The agent assigns a probability $p_{\pi}(a_t \mid s_t)$ to each possible action based on the current state $s_t$.

In our unified formulation of MDP for RAR, the LLM $\pi_\theta$ acts as the agent. 
The environment can take various forms, such as a knowledge repository~\cite{Yao23ReAct}. 
The set of possible actions in each state is $A = \{a_{\text{ret}}, a_{\text{ans}}\}$, where $a_{\text{ans}}$ denotes the halting action,  and $a_{\text{ret}}$ represents the retrieval action. 
Each intermediate state $s_t$ consists of a think $\tau_t$ followed by a search query $q_t$; i.e., $s_t = \langle \tau_t, q_t \rangle$. The final state $s_N$ contains a think $\tau_N$ followed by a final response $r$.
The transition probability $P(s_{t+1} \mid s_t, a_t)$ is determined by the LLM itself. An explicit reward function is not always required, as some agents operate without additional training.

\medskip
\noindent
\textbf{Consistency-based \MainTask.}
The core idea of consistency-based methods~\cite{xiong2023can, Hou24Decomposing} is to generate multiple responses for a given query by varying either the input prompt or the temperature parameter used in stochastic decoding~\cite{Korikov25Batched}. The idea builds upon the hypothesis of self-consistency theory, where correct reasoning processes, even if they are diverse, tend to have greater agreement in their final answer~\cite{Wang23SelfConsistency}.
The pairwise similarity among these responses is then computed and aggregated into a single confidence score~\cite{lin24bbox, bakman2025reconsidering}. 
Formally, consider a model $\pi$ parameterized with $\theta$, generates the most-likely response $r$ by setting the sampling temperature to less than one.
Then, $B$ additional responses $R_G = \{r^b\}_{b=1}^B$ are sampled using various sampling strategies such as increasing the temperature~\cite{Wang23SelfConsistency}, changing the input~\cite{Korikov25Batched}, or altering the reasoning path~\cite{qi2025rstar, li2025raspberry}.
A transformation function $\mathcal{\phi}$ is then applied to convert $R_G$ into a confidence score.
Our method builds on the reasoning path perturbation approach and employs \textit{Majority Voting}~\cite{Wang23SelfConsistency} as the transformation function, where confidence is the degree of consistency, measured by the proportion of sampled responses that match the most-likely response:
\begin{equation}
C(x, r) = \mathcal{\phi}(R_G, r) = \frac{1}{B} \sum_{b=1}^{B} \mathbb{I}(r^b \equiv r).
\label{eq:uncertianty_score}
\end{equation}

\section{Methodology}~\label{sec:methodology}

\setlength{\abovedisplayskip}{3pt} 
\setlength{\belowdisplayskip}{3pt} 

We propose \textbf{R}etrieval-Augmented \textbf{R}easoning \textbf{C}onsistency, \textbf{\proposedue}, to address \MainTaskAbbr\ in RAR models.
\proposedue\ is a consistency-based approach that performs \MainTaskAbbr\ in two main stages, as illustrated in Figure~\ref{fig:main_fig}: (i) generating the most-likely response, and (ii) sampling multiple generations.
The core idea of \proposedue\ is to perturb the reasoning paths of these multiple generations through a set of perturbation actions, denoted as $A^*$. In MDP terms, \proposedue\ temporarily replaces the main action set $A$ with $A^*$ for a single state, allowing the RAR model to interleave its generation process and explore new reasoning trajectories, queries, and documents.
We define three perturbation actions employed in \proposedue: (i) Query Paraphrasing, $a_{qp}$, (ii) Critical Rethinking, $a_{cr}$, and (iii) Answer Validation, $a_{av}$.
By employing a variety of actions to perturb the reasoning processes of LLMs, \proposedue\ captures both epistemic (model) uncertainty and aleatoric (data) uncertainty. This perspective is supported by \citet{liu2025uncertainty}, who argue that reasoning uncertainty encompasses both epistemic and aleatoric components. In this work, we do not seek to disentangle these two sources of uncertainty; rather, we measure their combined effect as total uncertainty.




In the following sections, we first formally describe how \proposedue\ perturbs the generation
path and then describe perturbation actions.


\subsection{\proposedue: Retrieval-Augmented Reasoning Consistency}
\proposedue\ models uncertainty for RAR as an MDP, in which multiple response generations are produced by temporarily replacing the action set $A$ with an alternative set $A^*$ at a randomly selected state $s_t$. 
First, the most-likely generation is produced as an MDP iteratively:
$$s_t \gets \pi_{\theta}\!\left(s_{t-1}, a_{t-1}\right); \quad t = 1, \dots, N,$$
where $N$ denotes the length of the reasoning path, determined by the agent $\pi_\theta$ when it selects the halting action $ a_{\text{ans}}$. The most-likely response $r$ is obtained from the final state $s_N$. For example, in the left part of Figure~\ref{fig:main_fig}, the most-likely response is “\textit{Orlando Magic},” derived from the most-likely generation highlighted in yellow. The middle part of the figure shows the multi-generation process where an action is randomly selected and applied to a state.

To construct the sampled response set $R_G$, we first fix the number of generations to $B$. In each generation, we uniformly sample an action $a^*$ from the perturbation action set $A^*$ and a perturbation state $s_t$ from the most-likely reasoning path $\{s_t\}_{t=1}^N$. 
An exception is the action $a_{av}$, for which the perturbation state is always set to $s_N$.
The agent then transitions from state $s_t$ to $s_{t+1}$ given action $a^*$:
$$s_{t+1} \gets \pi_{\theta}(s_t, a^*).$$
For the remainder of the path, the actions are sampled from the main action set $A$ until reaching a new end state $N^b$, determined by the agent:
$$s_{t+i+1} \gets \pi_{\theta}\!\left(s_{t+i}, a_{t+i}\right); \quad i = 1, \dots, N^b.$$
The final sampled response $r^b$ is obtained from the last state $s_{N^b}$ and added to $R_G$. 
After $B$ iterations, we obtain the response set $R_G = \{r^b\}_{b=1}^B$, on which the confidence score function (Eq.~\eqref{eq:uncertianty_score}) is applied to compute the final confidence score.
Algorithm~\ref{alg:r2c} provides a detailed description of the entire process in \proposedue.

\subsection{Perturbation Actions}~\label{sec:actions}

\noindent \textit{\textbf{A1: Query Paraphrasing (QP).}} 
Constructing effective search queries is critical for retrieving relevant documents; however, LLMs are not inherently optimized for this purpose~\cite{jiang2025s3, ma2023query, li2025raspberry}.
The QP action ($a_{qp}$) is introduced as a query optimization mechanism that enables the system to explore alternative semantic formulations of the original query. 
Precisely, when action $a_{qp}$ is applied, the think $\tau_t$ of the state $s_t$ is preserved and only the query changes.
Formally, the LLM $\pi_{\theta}$ takes action $a_{qp}$ at state $s_t$, transitioning to state $s_{t+1}$ with the same think $\tau_t$ and a new query $q_{t+1}$:
$$s_{t+1} =\langle \tau_t, q_{t+1} \rangle  \gets \pi_{\theta}( \langle \tau_t, q_t \rangle, a_{qp}).$$
Conceptually, QP perturbation tests whether the reasoning path is so fragile that paraphrasing the search query can alter its direction and lead to the retrieval of different documents. The QP action is implemented by prompting the LLM with a paraphrasing instruction.

\renewcommand{\arraystretch}{1.2}
\begin{table*}[t]
\centering 
\setlength{\tabcolsep}{0.9pt}
\caption{Performance of UQ methods measured by AUROC. In each column, the best and second-best methods are indicated by \textbf{bold} and \underline{underline}, respectively. Superscripts \textsuperscript{\dag} and \textsuperscript{\ddag} denote statistically significant differences according to the DeLong test ($p$ < 0.05), compared to \reabaseline\ and P(true), respectively, which are the two best-performing methods on average.}
\label{tab:uncertainty_estimation_evaluation}
\shrink
\begin{tabular}{l|ccc|ccc|ccc|ccc|ccc|c}
\hline

\textbf{RAG} &
\multicolumn{3}{c}{\textbf{SelfAsk~\cite{press2023selfask}}} &
\multicolumn{3}{c}{\textbf{ReAct~\cite{Yao23ReAct}}} &
\multicolumn{3}{c}{\textbf{Search-o1~\cite{li25searcho1}}} &
\multicolumn{3}{c}{\textbf{ReSearch~\cite{chen25research}}} &
\multicolumn{3}{c|}{\textbf{Search-R1~\cite{Jin25SearchR1}}} &
\multirow{2}{*}{\textbf{Avg.}} \\
\cline{0-15}
Uncer. M.  &
Popqa & Hotp. & Musi. &
Popqa & Hotp. & Musi. &
Popqa & Hotp. & Musi. &
Popqa & Hotp. & Musi. &
Popqa & Hotp. & Musi. & \\ \hline\hline
PE~\cite{Kadavath22PE} & 
55.34 & 59.11 & 48.61 & 36.75 & 39.95 & 41.14 & 39.86 & 53.44 & 51.59 & 40.93 & 61.03 & 56.69 & 48.49 & 48.27 & 50.90 & 48.81 \\

SE~\cite{Kuhn23SE} & 
64.26 & 68.01 & 54.68 & 49.73 & 40.37 & 41.69 & 67.56 & 65.84 & 54.61 & 53.88 & 64.72 & 63.43 & 59.11 & 53.45 & 61.38 & 57.51 \\

MARS~\cite{Bakman24mars} & 
54.70 & 59.48 & 51.53 & 41.29 & 40.42 & 41.97 & 43.28 & 57.69 & 56.80 & 40.03 & 61.76 & 57.58 & 48.33 & 49.28 & 49.94 & 50.27 \\

SAR~\cite{duan2024sar} & 
51.65 & 63.05 & 52.31 & 29.56 & 32.90 & 31.54 & 40.92 & 52.37 & 44.66 & 41.67 & 62.75 & 51.47 & 45.87 & 46.40 & 45.22 & 46.16 \\

LARS~\cite{yaldiz2025lars} & 
73.97 & 70.03 & 66.45 & 79.95 & 68.28 & 71.87 & 83.79 & 71.42 & 66.19 & 76.95 & 66.12 & 71.64 & 71.54 & 67.08 & 71.62 & 71.79 \\

NumSS~\cite{Kuhn23SE} & 
71.31 & 65.19 & 62.75 & 74.90 & 63.73 & 62.59 & 78.88 & 63.74 & 62.12 & 80.42 & 64.46 & 69.34 & 69.76 & 64.10 & 65.72 & 67.93 \\

EigV~\cite{lin24bbox}   & 
70.53 & 66.80 & 62.13 & 76.48 & 66.44 & 57.44 & 79.72 & 68.58 & 64.75 & 80.63 & 66.43 & 66.31 & 69.33 & 66.54 & 64.27 & 68.43 \\

ECC~\cite{lin24bbox}    & 
72.89 & 69.95 & 64.11 & 80.27 & 69.51 & 61.92 & 81.98 & 69.27 & 69.65 & 81.85 & 68.49 & 72.55 & 70.87 & 67.76 & 67.65 & 71.25 \\

Deg~\cite{lin24bbox}    & 
70.53 & 66.79 & 61.77 & 76.70 & 67.70 & 58.12 & 80.69 & 68.87 & 66.29 & 81.67 & 67.15 & 67.85 & 69.51 & 66.75 & 64.53 & 68.99 \\

\rrrbaseline~\cite{li2025raspberry}  & 
71.14 & 71.17 & \textbf{81.28} & 48.02 & 68.30 & \underline{75.99} & 65.95 & 73.87 & \underline{77.95} & 68.30 & 71.63 & 74.25 & 68.92 & 71.08 & 70.77 & 70.57 \\

\selfbaseline~\cite{Wang23SelfConsistency} & 
74.33 & 69.06 & 68.40 & 80.73 & 75.34 & 72.96 & \underline{81.26} & 72.34 & 76.87 & 82.01 & 72.14 & 77.89 & 71.63 & 69.04 & 70.36 & 74.29 \\

\reabaseline~\cite{qi2025rstar}  & 
\underline{77.02} & 70.90 & \underline{76.75} & \underline{81.53} & \underline{76.94} & 74.74 & 81.09 & \underline{74.97} & 77.01 & 82.50 & 75.22 & 77.86 & 73.22 & 72.79 & \underline{75.29} & 76.52 \\

P(true)~\cite{Kadavath22PE} & 
76.51 & \underline{77.66} & 78.65 & 75.07 & 73.45 & 71.51 & 78.57 & 73.19 & 74.83 & \underline{84.35} & \underline{77.77} & \underline{81.42} & \underline{75.73} & \underline{76.25} & 74.80 & \underline{76.65} \\

\textbf{\proposedue (our)}  & 
\textbf{80.08} &
\textbf{81.09}\textsuperscript{\dag} &
75.82 &
\textbf{84.75}\textsuperscript{\ddag} &
\textbf{83.25}\textsuperscript{\dag}\textsuperscript{\ddag} & 
\textbf{81.16}\textsuperscript{\dag}\textsuperscript{\ddag} & 
\textbf{87.09}\textsuperscript{\dag}\textsuperscript{\ddag} &
\textbf{79.66}\textsuperscript{\dag}\textsuperscript{\ddag} &
\textbf{83.22}\textsuperscript{\dag}\textsuperscript{\ddag} &
\textbf{86.02}\textsuperscript{\dag} &
\textbf{80.76}\textsuperscript{\dag} &
\textbf{82.39} &
\textbf{84.92}\textsuperscript{\dag}\textsuperscript{\ddag} &
\textbf{79.51}\textsuperscript{\dag} &
\textbf{80.08}\textsuperscript{\dag}\textsuperscript{\ddag} &
\textbf{81.99} \\

\hline\hline

\end{tabular}
\miniskip
\end{table*}

\medskip \noindent \textit{\textbf{A2: Critical Rethinking (CR).}} 
RAR models often suffer from the problem of \textit{self-criticism}, where they fail to recognize that previously retrieved information is noisy or irrelevant~\cite{Jin25SearchR1, jiang2025s3, Asai24selfrag, li2025raspberry}. Consequently, they continue to build their reasoning on top of earlier steps, even when those steps are uninformative and lack relevant content. This issue becomes particularly severe when it occurs in the early stages of the reasoning path.

The CR action ($a_{cr}$) critically reassesses the reasoning states produced up to state $s_t$. When applied at state $s_t$, it introduces a new state $s_{t+1}$ that the think $\tau_{t+1}$ explicitly evaluates the previously retrieved information as unhelpful, irrelevant, or misleading, and the accompanying search query $q_{t+1}$ is formulated to support this critical assessment.
Formally, in state $s_t$, the LLM $\pi_{\theta}$ is prompted with a critical rethinking instruction to generate a new state $s_{t+1}$:
$$s_{t+1} =\langle \tau_{t+1}, q_{t+1} \rangle  \gets \pi_{\theta}(\langle \tau_t, q_t \rangle, a_{cr}).$$
Conceptually, if the reasoning path so far has been incorrect, this action enables the system to adjust to a more reliable trajectory. If the path has been correct, CR strengthens its validity, thereby increasing confidence in the final outcome. 

\medskip \noindent \textit{\textbf{A3: Answer Validation (AV).}}
RAR models face difficulties validating their final response, detecting whether the generated response meets certain criteria of the query~\cite{Wang24FairEvaluators, Salemi25LiveRAG, He24cov, Jiang25RAGStar}.
One challenge is that the response is built upon a reasoning path that integrates both documents and the intermediate reasoning trajectory, which often leads the LLM to exhibit excessive confidence in its output. Another challenge is that different tasks and response types involve specific validation criteria, but RAR models are generally unaware of these requirements.

We introduce the AV action ($a_{av}$) to validate the final response by  prompting the LLM to reconsider its generation once a response has been produced, based on predefined criteria. Specifically, the LLM first generates a \textit{\textbf{query-aware reasoning path summary}}~\cite{shi2025refine, li25searcho1, Achkar25Summarize},
and then evaluates the final response using two criteria: (i) \textit{Groundedness}: is the response supported by the retrieved documents? and (ii) \textit{Correctness}: 
does the response appropriately and sufficiently address the query, given the available evidence?
%
Formally, let $D=\{D_1,\dots,D_{N-1}\}$ be the set of documents retrieved at states $[s_1, s_2, ...s_{N-1}]$. A model $\mathcal{M}$ generates the summary $\hat S$ of these documents: $\hat S = \mathcal{M}(x, D).$
The state $s_N$ is then updated with this summary, denoted as $\hat s_N$.
With the updated state $\hat s_N$,  the LLM $\pi_{\theta}$ is then instructed to generate a new state $s_{N+1}$:
$$ s_{N+1} \gets \pi_{\theta}(\hat s_N, a_{av}).$$
In principle, if the final response $r$ is validated as correct, the system outputs it directly in state $s_{N+1}$. Otherwise, if the validation indicates that the answer is incorrect or incomplete, the system begins a new reasoning path starting from $\langle \tau_{N+1}, q_{N+1}\rangle$.
\section{Experimental Setup}~\label{sec:experimental_setup}
\shrink


Our experiments consist of evaluation of uncertainty scores estimated by UQ methods as well as extrinsic evaluation on Abstention and Model Selection. In the following, we review our experimental setup for each of these tasks.

\subsection{Direct Evaluation of UQ Estimations}~\label{sec:exp_setup_uq}
\noindent
\textbf{\textit{Datasets}}:
We evaluate \proposedue\ on both single-hop and multi-hop QA tasks using the PopQA~\cite{mallen23popqa}, HotpotQA~\cite{Yang18HotpotQA}, and Musique~\cite{Trivedi22MuSiQue} datasets. Following prior work~\cite{trivedi2023ircot, jiang2023flare, Yao23ReAct, Moskvoretskii25Adaptive, bakman2025reconsidering}, we randomly sample $500$ queries from each dataset as the test set. We will release our sampled queries to improve reproducibility of our work. For the retrieval corpus, we use the 2018 Wikipedia dump~\cite{Karpukhin20Dense}.\footnote{\url{https://huggingface.co/datasets/PeterJinGo/wiki-18-corpus}} The number of retrieved documents is fixed to three across all models~\cite{mallen23popqa, Jin25SearchR1, Yao23ReAct, press2023selfask}.

\vspace{0.5em}
\noindent
\textbf{\textit{Evaluation Metrics}}:
To evaluate the quality of outputs, we follow \citet{Jin25SearchR1} and report the exact match, where a prediction is counted as correct if and only if it exactly matches one of the ground-truth responses. For evaluating \MainTaskAbbr\ methods, we follow prior work on UQ and use the threshold-free metric AUROC, which captures the correlation between uncertainty scores and response correctness~\cite{Kadavath22PE, Kuhn23SE, Bakman24mars}. As suggested by \citet{perez2025uncertainty}, significant differences between two AUROC values are assessed using the paired De Long test~\cite{delong1988comparing}.

\renewcommand{\arraystretch}{1.2}
\begin{table*}[t]
\shrink
\centering 
\setlength{\tabcolsep}{1.7pt}
\caption{Abstention performance measured by \textit{AbstainAccuracy} and \textit{AbstainF1} at a 0.9 confidence threshold. For each column, the best and second-best methods are indicated in bold and underlined, respectively. A superscript \textsuperscript{\dag} denotes a statistically significant difference compared to \reabaseline\ based on the McNemar test for Accuracy and the Bootstrap test for F1 ($p < 0.05$).}
\label{tab:abstention_performance}
\shrink
\begin{tabular}{l|ccc|ccc|ccc|ccc|ccc|c}
\hline

\textbf{RAG} &
\multicolumn{3}{c}{\textbf{SelfAsk~\cite{press2023selfask}}} &
\multicolumn{3}{c}{\textbf{ReAct~\cite{Yao23ReAct}}} &
\multicolumn{3}{c}{\textbf{Search-o1~\cite{li25searcho1}}} &
\multicolumn{3}{c}{\textbf{ReSearch~\cite{chen25research}}} &
\multicolumn{3}{c|}{\textbf{Search-R1~\cite{Jin25SearchR1}}} &
\multirow{2}{*}{\textbf{Avg.}} \\
\cline{0-15}
Uncer. M.  &
Popqa & Hotpot & Musiq. &
Popqa & Hotpot & Musiq. &
Popqa & Hotpot & Musiq. &
Popqa & Hotpot & Musiq. &
Popqa & Hotpot & Musiq. & \\ \hline\hline

\rowcolor{gray!20}
\multicolumn{17}{l}{\textit{Abstain Accuracy}} \\ 

\rrrbaseline~\cite{li2025raspberry} & 
61.4 & 65.6 & 73.0 & 60.8 & 72.2 & 82.4 & 56.6 & 66.2 & 68.2 & 59.2 & 64.4 & 64.0 & 60.6 & 64.0 & 63.6 & 65.48 \\

P(true)~\cite{Kadavath22PE} & 
\underline{70.6} & \underline{70.8} & \underline{80.4} & 70.4 & 67.2 & 75.0 & 72.6 & 65.8 & 75.0 & \underline{78.8} & \underline{71.4} & 78.2 & \underline{70.8} & \underline{70.0} & 74.4 & 72.76 \\ 

\selfbaseline~\cite{Wang23SelfConsistency} & 
68.6 & 64.2 & 76.2 & 74.0 & \underline{75.8} & 87.6 & 75.0 & 69.0 & \textbf{90.2} & 75.2 & 68.6 & \underline{83.6} & 65.4 & 62.8 & 77.8 & 74.27 \\

\reabaseline~\cite{qi2025rstar} &
69.2 & 64.6 & \underline{80.4} & \underline{74.4} & 75.4 & \underline{89.4} & \underline{75.6} & \underline{71.4} & 87.4 & 77.2 & 68.6 & 80.8 & 67.8 & 68.8 & \underline{80.6} & \underline{75.44} \\

\textbf{\proposedue\ (our)} & 
\textbf{77.2}\textsuperscript{\dag} &
\textbf{74.4}\textsuperscript{\dag} &
\textbf{88.6}\textsuperscript{\dag} &
\textbf{77.0} &
\textbf{76.8} &
\textbf{90.4} &
\textbf{80.4}\textsuperscript{\dag} &
\textbf{74.4} &
\underline{89.4} &
\textbf{81.6}\textsuperscript{\dag} &
\textbf{74.8}\textsuperscript{\dag} &
\textbf{84.0} &
\textbf{77.0}\textsuperscript{\dag} &
\textbf{73.4} &
\textbf{84.4}\textsuperscript{\dag} &
\textbf{80.25} \\

\rowcolor{gray!20}
\multicolumn{17}{l}{\textit{Abstain F1}} \\ 

\rrrbaseline~\cite{li2025raspberry} & 
62.52 & 70.54 & 82.75 & 71.59 & 80.82 & 89.69 & 56.33 & 71.50 & 78.99 & 54.86 & 63.37 & 73.68 & 55.12 & 61.20 & 73.92 & 69.79 \\

P(true)~\cite{Kadavath22PE} & 
\underline{73.60} & \underline{75.33} & 88.27 & 72.07 & 73.46 & 84.58 & 76.50 & 71.73 & 84.51 & 81.20 & 73.46 & 85.82 & \underline{71.03} & \underline{71.15} & 83.37 & 77.74 \\ 

\selfbaseline~\cite{Wang23SelfConsistency} & 
72.60 & 69.51 & 85.71 & \underline{79.10} & \underline{83.80} & 93.18 & 80.50 & 78.07 & \textbf{94.64} & \underline{81.71} & \underline{75.19} & \textbf{90.57} & 62.31 & 62.34 & 86.51 & 79.72 \\

\reabaseline~\cite{qi2025rstar} &
73.54 & 70.45 & \underline{88.41} & 79.01 & 83.17 & \underline{94.19} & \underline{81.24} & \underline{79.48} & 93.03 & 81.25 & 72.60 & 88.43 & 67.60 & 71.11 & \underline{88.27} & \underline{80.79} \\

\textbf{\proposedue\ (our)} & 
\textbf{82.35}\textsuperscript{\dag} &
\textbf{81.76}\textsuperscript{\dag} &
\textbf{93.77}\textsuperscript{\dag} &
\textbf{83.69}\textsuperscript{\dag} &
\textbf{85.08}\textsuperscript{\dag} &
\textbf{94.81} &
\textbf{86.42}\textsuperscript{\dag} &
\textbf{82.22}\textsuperscript{\dag} &
\underline{94.27}\textsuperscript{\dag} &
\textbf{84.40}\textsuperscript{\dag} &
\textbf{78.71}\textsuperscript{\dag} &
\underline{90.49}\textsuperscript{\dag} &
\textbf{80.80}\textsuperscript{\dag} &
\textbf{77.57}\textsuperscript{\dag} &
\textbf{90.97}\textsuperscript{\dag} &
\textbf{85.82} \\
\hline\hline

\end{tabular}
\miniskip
\end{table*}

\vspace{0.5em}
\noindent
\textbf{\textit{Models}}:
In line with prior work in RAR~\cite{Jin25SearchR1, chen25research}, we employ \textit{Qwen-2.5-7B-Instruct}~\cite{yang24Qwen25} as the generator LLM and path summary generator for action $a_{av}$. For \MainTaskAbbr, we sample $10$ responses per query with a temperature of $T=1.0$, while for correctness evaluation we generate the most-likely generation with $T=0.7$~\cite{soudani2025uncertainty, Bakman24mars}.
Retrieval is performed using a two-stage re-ranking pipeline: BM25~\cite{Robertson94bm25} is used for initial retrieval, followed by re-ranking with the pre-trained cross-encoder model \verb|ms-marco-MiniLM-L-6-v2| from the \verb|sentence-transformers| library.\footnote{\url{https://huggingface.co/cross-encoder/ms-marco-MiniLM-L6-v2}} All experiments are conducted on four Nvidia A100 GPUs, each with 40 GB memory, requiring $\sim$1500 GPU hours in total. 

\vspace{0.5em}
\noindent
\textbf{\textit{Baselines}}:
We use two sets of baselines: 
(1) \textbf{\textit{Path-based}} methods, which focus on generating multiple responses based on diverse reasoning paths and differ mainly in how they initiate new generations relative to the most-likely generation.
Self-Consistency (\selfbaseline)~\cite{Wang23SelfConsistency} ignores the most-likely generation and instead produces a diverse set of reasoning paths, starting from scratch. 
Reasoning Consistency (\reabaseline)~\cite{qi2025rstar} randomly truncates the most-likely reasoning path at different random steps and regenerates the response based on the subsequent reasoning steps. 
Retrieval-Retained Reasoning Consistency (\rrrbaseline)~\cite{li2025raspberry} applies truncation only after the last retrieved document. 
In all cases, the generated responses are aggregated into a final score using majority voting.
2) \textbf{\textit{Estimation-based}} methods include both white-box and black-box approaches. 
The white-box methods are PE~\cite{Kadavath22PE}, SE~\cite{Kuhn23SE}, MARS~\cite{Bakman24mars}, LARS~\cite{yaldiz2025lars}, and SAR~\cite{duan2024sar}. The black-box methods are NumSS~\cite{Kuhn23SE}, EigV, ECC, Deg~\cite{lin24bbox}, and P(true)~\cite{Kadavath22PE}. 
All of these methods rely on generations obtained in the same way as \selfbaseline.
All \MainTaskAbbr\ methods are implemented using TruthTorchLM~\cite{yaldiz2025truthtorchlm}.

\subsection{Extrinsic UQ Evaluation via Abstention}~\label{sec:abstention}
\noindent
\textbf{\textit{Task Formulation.}}
Declining to respond due to uncertainty is an important application of \MainTaskAbbr~\cite{Feng24Hallucinate}.
An abstention function determines whether the model should withhold an answer.
In our setup, this decision is guided by confidence scores. We introduce a threshold $\tau_{\text{abs}}$: if the confidence is less than $\tau_{\text{abs}}$, the model abstains; otherwise, it produces an answer.
Formally, given a response $r$ to a query $x$ and a confidence function $C$, we define the abstention function $f_{\text{abs}}$ as:
$$
f_{\text{abs}}(r) =
\begin{cases}
\text{true}, & \text{if } C(x, r) <\tau_{\text{abs}} \\
\text{false}, & \text{otherwise}.
\end{cases}
$$

\vspace{0.5em}
\noindent
\textbf{\textit{Baselines \& Evaluation Metrics.}}
As baselines, we implement the abstention task using confidence scores derived from different \MainTaskAbbr\ methods, including P(true), \rrrbaseline, \reabaseline, \selfbaseline, and \proposedue. 
For evaluation, we follow \citet{Feng24Hallucinate} and report two metrics: \textit{AbstainAccuracy} and \textit{AbstainF1}. These metrics assess how well a model balances answer correctness with appropriate abstention.
We define a confusion matrix with four outcomes: (A) answered correctly, (B) abstained correctly, (C) answered incorrectly, and (D) abstained incorrectly. 
Based on this matrix, \textit{AbstainAccuracy} is defined as $\frac{A + D}{A + B + C + D}$ and measures whether abstention decisions are correct overall. Ideally, a model should abstain when it would otherwise answer incorrectly, and provide an answer when it is correct.
\textit{AbstainF1} captures the trade-off between reliability and answer coverage. It is computed as the harmonic mean of precision ($\frac{D}{B + D}$) and recall ($\frac{D}{C + D}$). This metric penalizes both unnecessary abstentions and incorrect answers, providing a balanced evaluation of abstention behavior.
%
The threshold $\tau_{\text{abs}}$ for each UQ method is selected via a sweep over the validation sets. 

\subsection{Extrinsic UQ Evaluation via Model Selection}\label{sec:model_selection}

\noindent\textbf{\textit{Task Formulation.}}
Model selection (or selection-based model ensemble \cite{Chen25Harnessing, Jiang23LLMBlender}) aims to select a final response for a question based on multiple candidate responses generated by different systems~\cite{Chen25Harnessing}.
Formally, given a user query $x$ and a set of systems $\{S_1, S_2, \dots, S_M\}$, each system $S_i$ produces a response candidate $r_i$.  
A model selection method $\mathcal{M}(x, R)$ then selects the final response $\hat r$ from all candidate responses $R = \{r_1, r_2, \dots, r_M\}$. 

\vspace{0.5em}
\noindent
\textbf{\textit{\proposedue\ Select}.}
The proposed \proposedue\ Select utilizes the confidence scores derived from \proposedue. Our method groups semantically similar responses into $K$ clusters.
Following~\cite{li2025raspberry, yaldiz2025truthtorchlm}, we use \textit{Qwen-2.5-7B-Instruct} to compute pairwise semantic similarities between candidate responses and cluster similar ones.
We then assign a confidence score to each cluster $g_i$:  
$$R_{G} = \{\langle r_{_{1}}, c_{g_1} \rangle, \langle r_{_{2}}, c_{g_1} \rangle, \dots, \langle r_{_{K}}, c_{g_K} \rangle \}.$$
The confidence score of each cluster $g=\{r_i\}_{i=1}^m$ is computed by aggregating the confidence scores of its members: $c_g = \sum_{r_i \in g} C(x, r_i),$ where the $C$ function provides the confidence score for the response $r_i$.
The final response $\hat r$ is the response with the highest confidence score: $\hat r = \arg\max_{j} c_{g_j}.$
If no clustering is applied, each response constitutes a cluster and $M=K$; we refer to this variation as \emph{\proposedue\ Select w/o clustering}.

\renewcommand{\arraystretch}{1.0}
\begin{table}[t]
\centering
\setlength{\tabcolsep}{2.1pt}
\caption{Model Selection performance measured by exact match. The superscript \textsuperscript{\dag} denotes a statistically significant difference from the best-performing baseline (underlined), according to the Wilcoxon test ($p$ < 0.05).}
\label{tab:model_seletion_res}
\shrink 
\begin{tabular}{l|cccc}
\hline
\textbf{RAG System} & \textbf{PopQA} & \textbf{HotpotQA} & \textbf{Musique} & \textbf{Average}\\
\hline\hline
\rowcolor{gray!20}
\multicolumn{5}{l}{\textit{Vanilla LLM \& RAG}} \\
Direct                          & 18.8 & 20.8 & 2.6 & 14.1 \\
CoT                             & 17.6 & 22.2 & 5.8 & 15.2 \\
Vanilla RAG                     & 30.2 & 18.6 & 4.4 & 17.7 \\
IRCoT~\cite{trivedi2023ircot}   & 34.6 & 27.2 & 5.4 & 22.4 \\
FLARE~\cite{jiang2023flare}     & 31.6 & 25.2 & 8.4 & 21.7 \\
DRAGIN~\cite{su2024dragin}      & 28.6 & 23.2 & 4.6 & 18.8 \\
\rowcolor{gray!20}
\multicolumn{5}{l}{\textit{Retrieval-Augmented Reasoning (RAR)}} \\
SelfAsk~\cite{press2023selfask} & 35.6 & 33.0 & 10.4 & 26.3 \\
ReAct~\cite{Yao23ReAct}         & 36.8 & 27.8 & 10.8 & 25.1 \\
Search-o1~\cite{li25searcho1}   & 33.2 & 29.0 & 10.0 & 24.1 \\
ReSearch~\cite{chen25research}  & 38.6 & 38.8 & 16.6 & 31.3 \\
Search-R1~\cite{Jin25SearchR1}  & 41.6 & 41.4 & 16.0 & 33.0 \\
\rowcolor{gray!20}
\multicolumn{5}{l}{\textit{Model Selection RAR}} \\
Random                               & 31.4 & 31.6 & 10.6 & 24.5 \\
LLMBlender~\cite{Jiang23LLMBlender}  & 34.4 & 36.0 & 12.0 & 27.5 \\
- \textit{w/o clustering}            & 35.6 & 31.8 & 9.6  & 25.7 \\
- \textit{w/o clus. \& conf.}        & 32.0 & 26.0 & 8.6  & 22.2 \\ 
RAGEnsemble~\cite{chen2025revisiting}& \underline{45.6} & \underline{46.0} & \underline{19.4} & \underline{37.0} \\
- \textit{w/o clustering}            & 44.4 & 40.8 & 18.0 & 34.4 \\
- \textit{w/o clus. \& conf.}        & 43.2 & 37.2 & 13.0 & 31.1 \\
\textbf{\proposedue Select (our)}    & \textbf{46.8\textsuperscript{\dag}} & \textbf{50.2\textsuperscript{\dag}} & \textbf{22.6\textsuperscript{\dag}}  &  \textbf{39.9}\\
- \textit{w/o clustering}            & 45.4 & 44.0 & 19.2 & 36.2 \\
\hdashline
Ideal Model Selection                & 55.0 & 57.2 & 30.0 & 47.4 \\
\hline \hline
\end{tabular}
\end{table}

\vspace{0.5em}
\noindent
\textbf{\textit{Baselines \& Evaluation Metrics.}}
We evaluate \proposedue\ Select performance against both single RAG and RAR systems, as well as existing selection-based ensemble approaches.  
As baselines, we consider LLMBlender~\cite{Jiang23LLMBlender} and RAGEnsemble~\cite{chen2025revisiting}. LLMBlender is a trained reward model that given a user query and a set of candidate responses, ranks the responses accordingly. RAGEnsemble is an instruction-based approach that selects a single final answer from a set of candidate responses. 
For all selection methods, including the baselines and \proposedue\ Select, the response candidates are obtained from  SelfAsk~\cite{press2023selfask}, ReAct~\cite{Yao23ReAct}, Search-o1~\cite{li25searcho1}, ReSearch~\cite{chen25research}, and Search-R1~\cite{Jin25SearchR1}.
For evaluation, we report the correctness of the final answer using exact match.

\section{Results}~\label{sec:results}

\setlength{\abovedisplayskip}{2pt} 
\setlength{\belowdisplayskip}{2pt} 

\shrink
We present a set of experiments that address the following research questions:
\textbf{RQ1}: 
How does \proposedue\ perform in quantifying uncertainty for different RAR models? (Sec.~\ref{sec:uq_performance}), 
\textbf{RQ2}: 
How does \proposedue\ perform as an external signal on downstream tasks, such as Abstention and Model Selection? (Sec.~\ref{sec:task_eval}), 
\textbf{RQ3}: What factors contribute to the effectiveness of \proposedue? (Sec.~\ref{sec:rq3}),
\textbf{RQ4}: How does \proposedue\ balance effectiveness and efficiency? (Sec.~\ref{sec:rq4}), 
\textbf{RQ5}: What is the effect of different actions in the performance of \proposedue? (Sec.~\ref{sec:action_selection}).
\textbf{RQ6}: Does \proposedue\ generalize across different backbone LLMs? (Sec.~\ref{sec:other_llms})


\subsection{\MainTask\ Performance}~\label{sec:uq_performance}
\emph{\textbf{RQ1}} evaluates the performance of \proposedue\ compared to other \MainTaskAbbr\ methods. Table~\ref{tab:uncertainty_estimation_evaluation} presents results on five RAR systems across three datasets. The findings indicate that most white-box methods, i.e., PE~\cite{Kadavath22PE}, SE~\cite{Kuhn23SE}, MARS~\cite{Bakman24mars}, and SAR~\cite{duan2024sar}, perform relatively poorly, with AUROC values ranging from about 30 to 60. This weakness stems from their overreliance on token probabilities. 
In contrast, black-box methods, such as NumSS~\cite{Kuhn23SE}, EigV, ECC, and Deg~\cite{lin24bbox}, generally outperform white-box methods, reaching average AUROC values between roughly 60 and 80. Interestingly, P(true) ranks as the second-best method in terms of average AUROC across all approaches, highlighting that black-box methods become much more effective when using an LLM as the scoring function. This advantage largely comes from their stronger reliance on textual diversity. 
The supervised method LARS, optimized for QA with no reasoning,  achieves AUROC scores between 65 and 85 across all cases, surpassing both white-box and black-box approaches on average. 
This finding highlights the potential of supervised \MainTaskAbbr\ as a promising direction for future research.

\begin{figure}[t]
  \centering
  \includegraphics[width=0.46\textwidth]{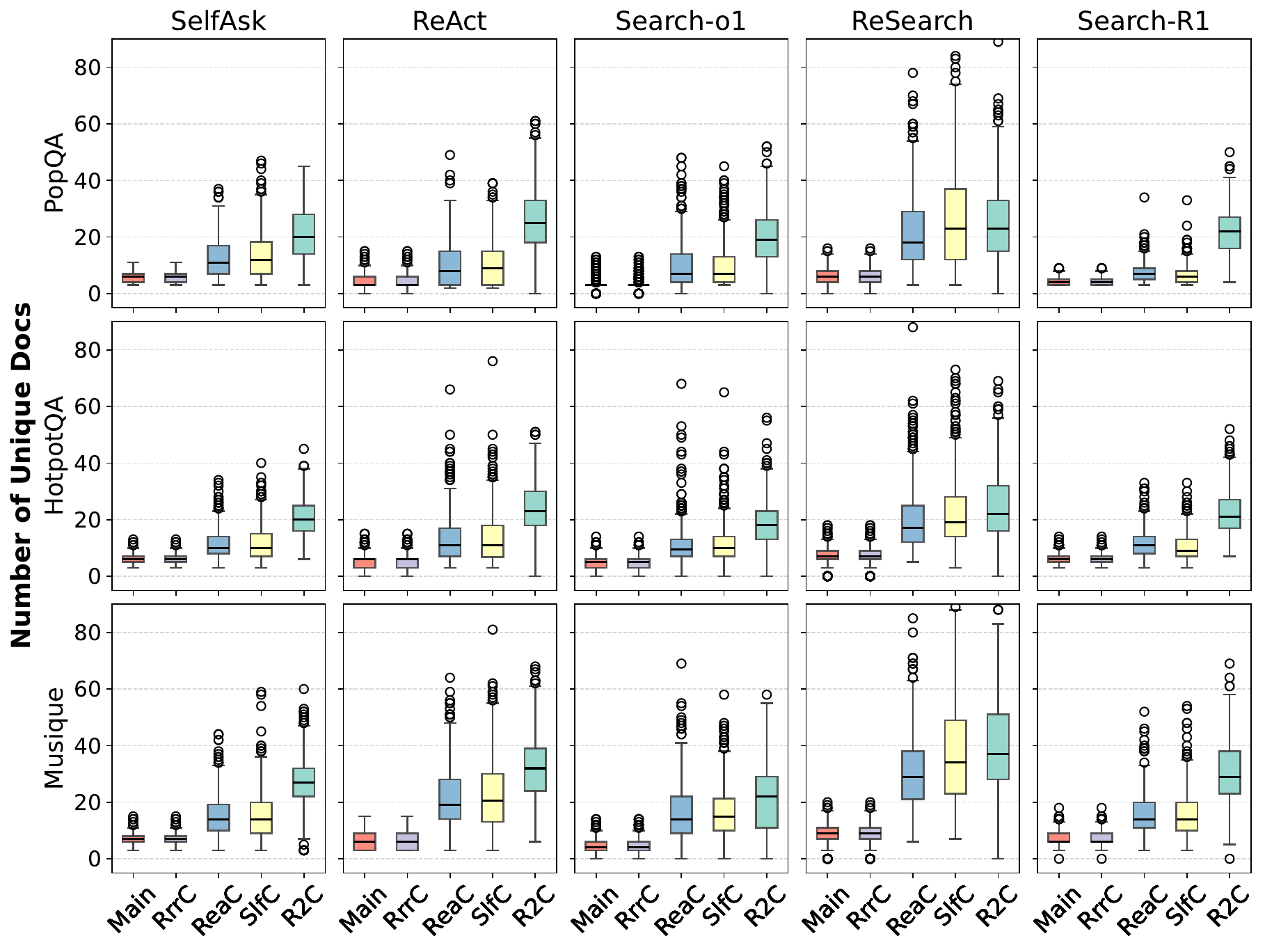}
  \shrink
  \caption{Distribution of the number of unique retrieved documents for the most-likely path (main) and multi-generations.}
  \label{fig:num_unique_retrieved_docs}
  \shrink
\end{figure}

\begin{figure}[t]
  \centering
  \includegraphics[width=0.46\textwidth]{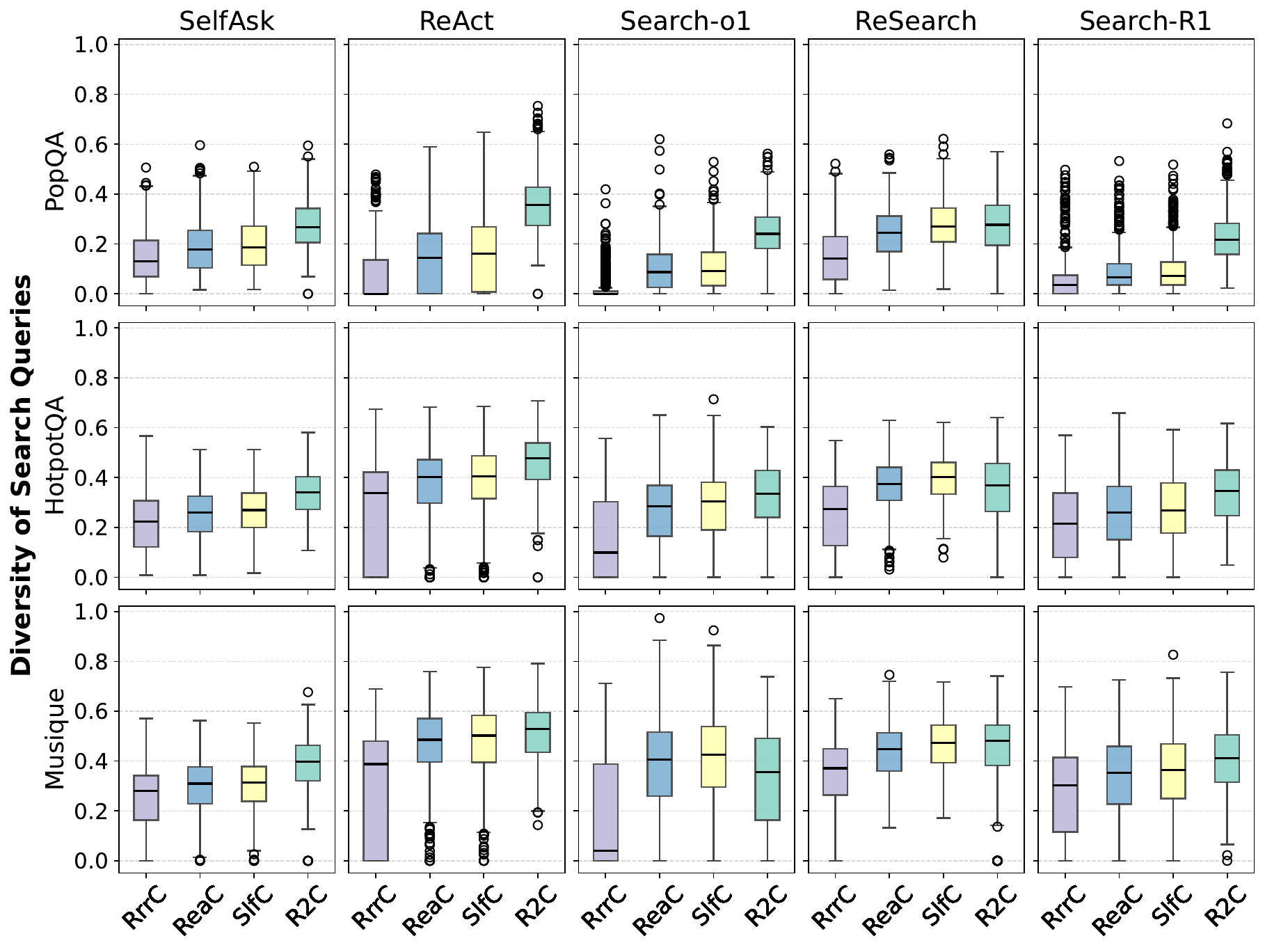}
  \shrink
  \caption{Distribution of diversity scores for search queries generated for each user query across reasoning paths.}
  \label{fig:search_queries_diversity}
  \shrink
\end{figure}

Path-based methods are another group of approaches that we evaluate, including the proposed \proposedue.
\rrrbaseline\ achieves an average AUROC of 70.57, which is the lowest among these methods. This indicates that simply keeping the documents in the reasoning path and regenerating the last state is not effective.
\selfbaseline\ and \reabaseline\ perform roughly on par, and
\reabaseline\ outperforms \selfbaseline\ on average, showing that regenerating from the top of the reasoning path does not necessarily guarantee a better uncertainty score.
Finally, \textbf{\proposedue}\ outperforms all methods by a large margin, with an absolute improvement of 5\% on average. These findings suggest that applying actions to the reasoning path enables exploration of a wider variety of possible reasoning states, leading to more reliable scores.

\begin{figure}[t]
  \centering
  \includegraphics[width=0.46\textwidth]{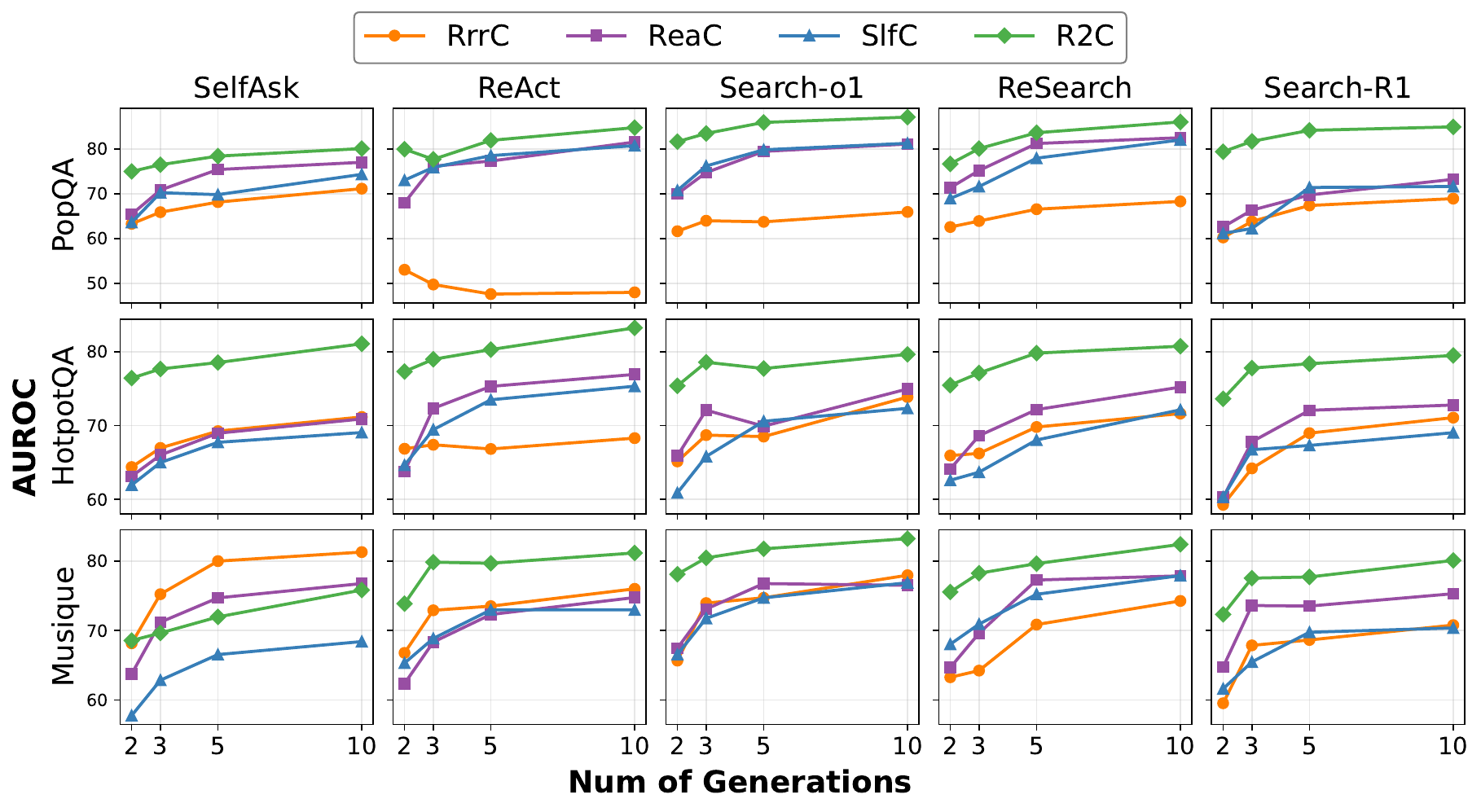}
  \shrink
  \caption{Performance of \MainTaskAbbr\ methods with varying numbers of generations.}
  \label{fig:num_generations}
  \shrink
\end{figure}

\begin{figure}[t]
  \centering
  \includegraphics[width=0.46\textwidth]{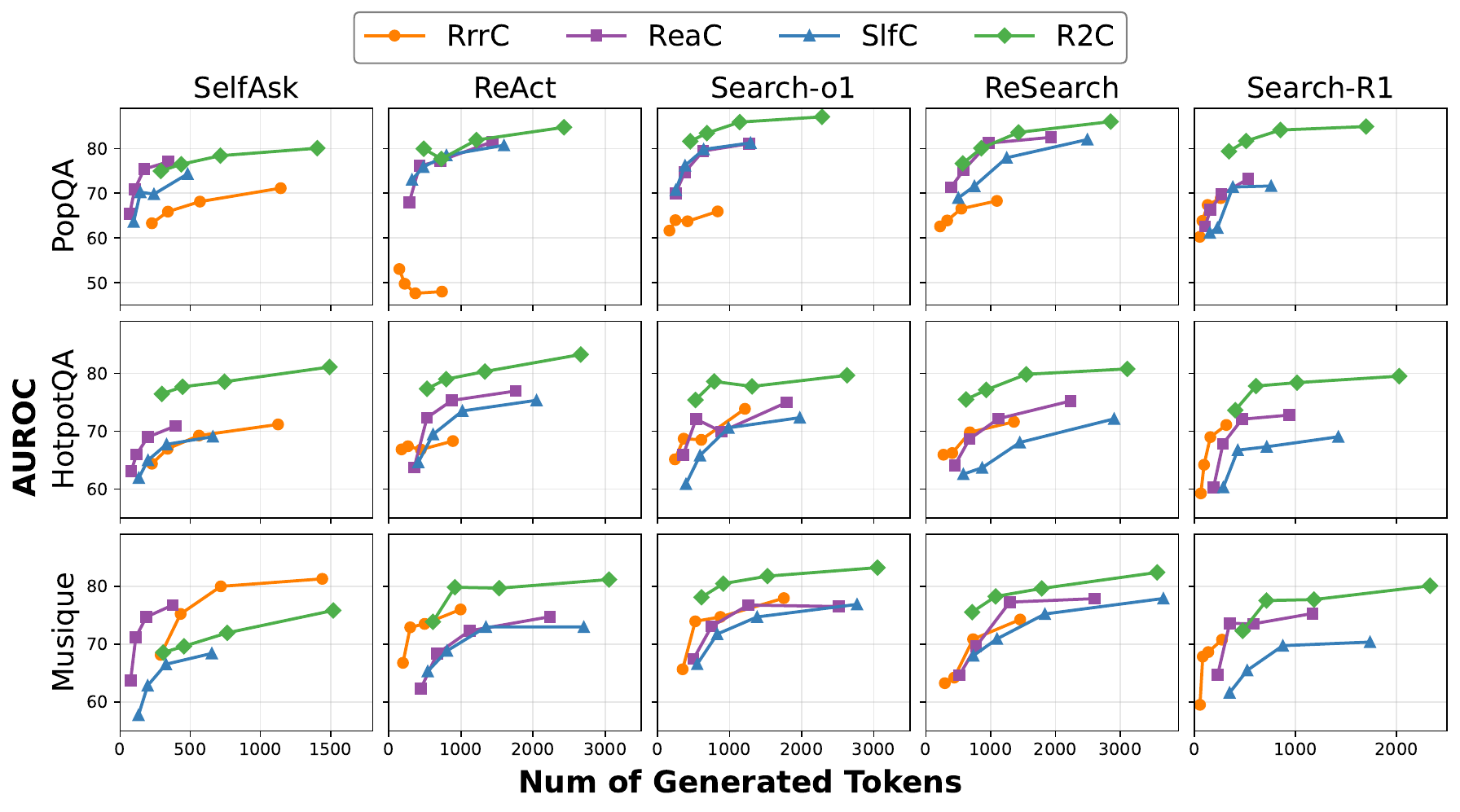}
  \shrink
  \caption{Performance of \MainTaskAbbr\ methods with varying number of generated tokens, illustrating the trade-off between effectiveness and efficiency.}
  \label{fig:auroc_vs_ntokens}
  \shrink
\end{figure}

\subsection{Extrinsic Evaluation Results}~\label{sec:task_eval}
\emph{\textbf{RQ2}} evaluates the performance of \proposedue\ in Abstention and Model Selection.
For Abstention, Table~\ref{tab:abstention_performance} shows the performance on RAR models and datasets. 
Abstain Accuracy measures whether both  abstentions and non-abstentions are detected correctly, while AbstainF1 captures the effectiveness of identifying cases where the model should abstain.
Considering Abstain Accuracy, \proposedue\ outperforms the baselines in all cases except on the Musique dataset, where it performs on par with the second-best baseline in ReAct, Search-o1, and ReSearch.
Considering Abstain F1, \proposedue\ significantly outperforms the best baseline in all setups, expect for ReAct on Musique. This indicates that \proposedue\ scores are more reliable at identifying cases where abstention is appropriate. 
Overall, compared to other \MainTaskAbbr\ methods, \proposedue\ achieves significantly better performance in most cases—on average, about 5\% higher than the second-best model.
These results demonstrate that the uncertainty scores generated by \proposedue\ are reliable enough for the system to decide when to refrain from answering. 

For model selection, Table~\ref{tab:model_seletion_res} reports the results of individual RAG and RAR systems, as well as selection-based ensemble models. 
Among individual systems, RAR models outperform RAG models for both simpler single-hop questions (PopQA) and more complex multi-hop queries (HotpotQA and Musique).
%
Within selection-based ensemble models, LLMBLender performs poorly, even worse than fine-tuned single RAR models such as ReSearch and Search-R1. In contrast, RAGEnsemble surpasses both LLMBLender and all single RAR models, highlighting the advantage of an instruction-tuned model for selecting the final response.
Finally, \textbf{\proposedue\ Select} achieves the best performance, significantly outperforming all RAR and RAG systems, with  3.7\% average improvement on HotpotQA and Musique datasets. 
\textbf{These results confirm that the \proposedue\ score is a reliable criterion, not only within a single system but also across different systems.}
Interestingly, using \proposedue\ confidence scores alone, without clustering, improves selection-based ensemble models in comparable settings, highlighting the informativeness of \proposedue\ scores 
even for models that are not trained on confidence.

\begin{figure}[t]
  \centering
  \includegraphics[width=0.44\textwidth]{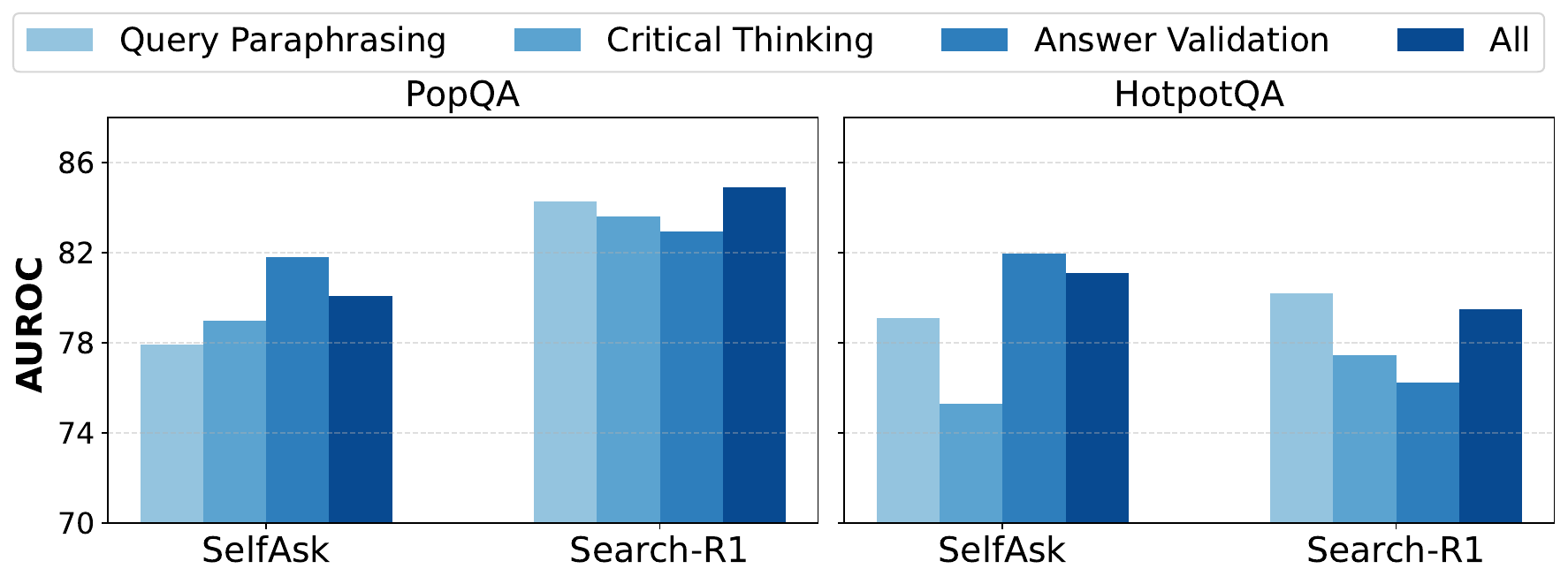}
  \caption{Performance of \proposedue\ with different action sets. 
  }
  \label{fig:action_set}
  \shrink
\end{figure}

\renewcommand{\arraystretch}{1.4}
\begin{table*}[t]
\centering 
\footnotesize
\setlength{\tabcolsep}{1.3pt}
\caption{Performance of UQ methods measured by AUROC across different LLMs. For each LLM and for each column, the best and second-best methods are indicated by \textbf{bold} and \underline{underline}, respectively. Superscripts \textsuperscript{\dag} and \textsuperscript{\ddag} denote statistically significant differences according to the DeLong test ($p < 0.05$), compared to \selfbaseline\ and \reabaseline, respectively, which are the two best-performing methods on average.}
\label{tab:other_llms}
\shrink
\begin{tabular}{l|ccc|ccc|ccc|c|ccc|ccc|ccc|c}
\hline

& \multicolumn{10}{c|}{\textbf{Gemma2-9b-it}} & \multicolumn{10}{c}{\textbf{GPT-4o-mini}} \\\hline
RAG &
\multicolumn{3}{c}{SelfAsk~\cite{press2023selfask}} &
\multicolumn{3}{c}{ReAct~\cite{Yao23ReAct}} &
\multicolumn{3}{c|}{Search-o1~\cite{li25searcho1}} &
\multirow{2}{*}{Avg.} & 
\multicolumn{3}{c}{SelfAsk~\cite{press2023selfask}} &
\multicolumn{3}{c}{ReAct~\cite{Yao23ReAct}} &
\multicolumn{3}{c|}{Search-o1~\cite{li25searcho1}} &
\multirow{2}{*}{Avg.}\\
\cline{1-10}\cline{12-20}
Uncer. M.  &
Popqa & Hotpot & Musiq. &
Popqa & Hotpot & Musiq. &
Popqa & Hotpot & Musiq. & &
Popqa & Hotpot & Musiq. &
Popqa & Hotpot & Musiq. & 
Popqa & Hotpot & Musiq. & \\ \hline\hline

\rrrbaseline~\cite{li2025raspberry} &
76.64 & 71.01 & 76.16 & 57.74 & 72.93 & 75.32 & 55.72 & 63.96 & 64.33 & 68.20 &
63.07 & \underline{70.96} & \underline{83.69} & 53.21 & 58.16 & 54.78 & 70.49 & \underline{77.27} & 84.04 & 68.41 \\

P(true)~\cite{Kadavath22PE} &
71.33 & 74.90 & 73.44 & 76.56 & 73.08 & 75.26 & 76.32 & 72.45 & 74.62 & 74.22 &
--- & --- & --- & --- & --- & --- & --- & --- & --- & --- \\

\selfbaseline~\cite{Wang23SelfConsistency} &
\underline{76.82} & 72.70 & 73.59 & \underline{81.79} & \underline{74.34} & 76.84 & \underline{79.17} & 70.59 & \underline{76.07} & 75.77 &
\underline{76.59} & 58.28 & 54.04 & 81.08 & 73.32 & 75.83 & 78.52 & 74.39 & 81.83 & 72.65 \\

\reabaseline~\cite{qi2025rstar} &
76.24 & \underline{79.77} & \underline{78.73} & 80.16 & 71.33 & \textbf{80.80} & 78.45 & \underline{73.61} & 75.77 & \underline{77.21} & 
75.53 & 60.58 & 54.66 & \underline{81.53} & \underline{74.34} & \underline{81.10} & \underline{79.41} & 75.09 & \underline{85.09} & \underline{74.15} \\

\textbf{\proposedue\ (our)} &
\textbf{80.29}\textsuperscript{\ddag} &
\textbf{81.86} &
\textbf{80.31} &
\textbf{82.87} &
\textbf{80.62}\textsuperscript{\dag}\textsuperscript{\ddag} &
\underline{80.37} &
\textbf{87.65}\textsuperscript{\dag}\textsuperscript{\ddag} &
\textbf{76.95}\textsuperscript{\dag}\textsuperscript{\ddag} &
\textbf{77.39} &
\textbf{80.92} &

\textbf{86.56}\textsuperscript{\dag}\textsuperscript{\ddag} &
\textbf{87.82}\textsuperscript{\dag}\textsuperscript{\ddag} &
\textbf{84.28}\textsuperscript{\dag}\textsuperscript{\ddag} &
\textbf{81.56} &
\textbf{75.84} &
\textbf{81.94}\textsuperscript{\dag} &
\textbf{82.48}\textsuperscript{\dag} &
\textbf{84.59}\textsuperscript{\dag}\textsuperscript{\ddag} &
\textbf{86.29}\textsuperscript{\dag} &
\textbf{83.48} \\
\hline\hline
\end{tabular}
\end{table*}

\subsection{Explaining the Strength of \proposedue}~\label{sec:rq3}
\emph{\textbf{RQ3}} examines why \proposedue\ outperforms other \MainTaskAbbr\ methods.
To this end, we analyze the diversity of reasoning paths as an indicator of how effectively \proposedue\ captures the uncertainty arising from both the generator and the retriever, i.e., the two key sources of uncertainty in our framework.
We report the number of unique documents retrieved during the multi-generation step, as shown in Figure~\ref{fig:num_unique_retrieved_docs}. On average, \proposedue\ retrieves 24.71 unique documents for a single confidence score, whereas \rrrbaseline, \selfbaseline, and \reabaseline\ retrieve 5.81, 15.35, and 16.4 documents, respectively. 

We further report on the diversity of generated search queries, as a proxy for the LLM's thinking process. Query diversity is measured based on the average pair-wise similarity of generated queries~\cite{cox2021directed, zhang2025evaluating}. Formally, given $n$ search queries $\{q_1, \dots, q_n\}$, query diversity is computed as:  
$$\text{Query Diversity} = 1 - \frac{2}{n(n - 1)} \sum_{i=1}^n\sum_{i < j}^n cos(e_i,e_j),$$
where $cos(.)$ represents cosine similarity function, and $e_i$ denotes the normalized embedding of query $q_i$ (i.e., $|e_i| = 1$) obtained from \verb|sentence-transformers/all-MiniLM-L6-v2|.
Figure~\ref{fig:search_queries_diversity} presents the results.
On average, \proposedue\ achieves a diversity score of $0.35$, while \rrrbaseline, \selfbaseline, and \reabaseline\ achieve $0.20$, $0.28$, and $0.30$, respectively. These results indicate that \proposedue\ generates more diverse search queries when estimating confidence.
These findings indicate that \proposedue\ effectively captures the uncertainty of both the retriever and the generator by sequentially diversifying their inputs.


\subsection{Effectiveness vs. Efficiency}~\label{sec:rq4}
\emph{\textbf{RQ4}} explores the trade-off between the effectiveness and efficiency of \proposedue.
Figure~\ref{fig:num_generations} presents the relationship between AUROC and the number of response generations across various datasets and  RAR models.
The results demonstrate that \proposedue\ consistently outperforms other methods, even with a smaller number of generations.
On average, across all datasets and RAR models, \proposedue\ achieves an AUROC performance of about 77\% with only three generations, which is comparable to the performance of \selfbaseline\ and \reabaseline, requiring 10 generations to reach a similar level.

For a deeper exploration of the trade-offs between effectiveness and efficiency, we measure efficiency using the total number of generated tokens, following prior work~\cite{perez2025uncertainty}. Figure~\ref{fig:auroc_vs_ntokens} reports AUROC performance under equal token-generation budgets. 
It shows \proposedue\ achieves higher AUROC scores than other methods given the same number of generated tokens for multi-hop datasets. For the single-hop dataset, PopQA, \proposedue\ performs comparably to SelfAsk, ReAct, and ReSearch, but surpasses them on Search-o1 and Search-R1. 
 Moreover, \proposedue\ produces on average around 700 tokens with three generations, reaching the same AUROC score as the baselines with approximately 1,700 tokens with 10 generations. These findings demonstrate that \proposedue\ improves efficiency by roughly 2.5 times.
Overall, these results indicate that \proposedue\ outperforms other \MainTaskAbbr\ methods in terms of both effectiveness and token generation efficiency.
Beyond improving token-generation efficiency, \proposedue\ is well suited for real-world deployment because it enables parallel generation of fully independent responses, thereby addressing the typical latency concerns associated with sampling-based methods. In practice, the confidence score can be computed at the cost of at most one additional generation beyond the most-likely response.


\subsection{Action Selection}~\label{sec:action_selection}
\emph{\textbf{RQ5}} investigates the role of the action set in \proposedue. 
Figure~\ref{fig:action_set} illustrates the performance of \proposedue\ with different action sets. In the first three bars, only a single action is used, meaning that in each generation, the cut point in the reasoning path is chosen randomly, while the action itself remains fixed. The fourth bar represents our main setup, where both the action and the cut point are selected randomly.
For Search-R1 and PopQA, the main setup outperforms all other configurations, whereas in HotpotQA, QP performs slightly better. For SelfAsk,  the AV action achieves the best performance, with the main setup ranking second. These results indicate that while our main setup is generally robust, there are still potentials to design action configurations better suited to specific RAR systems.
To further improve \proposedue, future work could focus on learning or adaptively selecting perturbations (e.g., via reinforcement learning), which we expect would enhance both effectiveness and efficiency.

\subsection{Generalization Across Backbone LLMs}~\label{sec:other_llms}
\emph{\textbf{RQ6}} investigates the robustness of \proposedue\ to changes in the generation model. Table~\ref{tab:other_llms} summarizes the performance of \MainTaskAbbr\ methods under alternative LLM backbones.
To ensure diversity, we evaluate one open-source LLM, \textit{Gemma2-9b-it},\footnote{\url{https://huggingface.co/google/gemma-2-9b-it}} and one closed-source LLM, \textit{GPT-4o-mini}. We do not evaluate ReSearch and Search-R1 in this setting, as their fine-tuned models are only available for \textit{Qwen} family.
With \textit{Gemma2-9b-it} as the backbone, \proposedue\ outperforms the second-best method, \reabaseline, by more than 3\%, indicating that the effectiveness of \proposedue\ is largely independent of the choice of LLM.

For \textit{GPT-4o-mini}, \proposedue\ outperforms \reabaseline\ by more than 9\%. The P(true) metric cannot be reported in this setting because token-level probabilities are not accessible. 
In the case of SelfAsk, we observe a different performance ordering, in which \rrrbaseline\ emerges as the second-best method after \proposedue, most notably on the Musique dataset. This shift is primarily due to the reduced effectiveness of SelfAsk when \textit{GPT-4o-mini} is used as the backbone compared to \textit{Qwen-2.5-7B-Instruct}, despite additional prompt engineering. In particular, SelfAsk achieves exact match accuracies of 18.6\%, 12.2\%, and 2.0\% on PopQA, HotpotQA, and Musique, respectively, which are lower than the corresponding results reported in Table~\ref{tab:model_seletion_res}.
Overall, \proposedue\ consistently achieves the highest average performance by a substantial margin, demonstrating strong generalization across different LLM backbones.

\section{Conclusions and Future Work}~\label{sec:conclusions}

This paper introduces a novel and theoretically grounded \MainTaskAbbr\ method for retrieval-augmented reasoning (RAR) systems, called Retrieval-Augmented Reasoning Consistency (\proposedue). We argue that an effective \MainTaskAbbr\ method should account for different  sources of uncertainty and accurately reflect them in its final score. \proposedue\ models uncertainty stemming from both the retriever and the generator by perturbing the reasoning process through a series of actions, including query paraphrasing, critical rethinking, and answer validation.
Comprehensive experiments conducted on three datasets and five RAR models demonstrate that \proposedue\ improves AUROC by more than 5\% on average compared to state-of-the-art \MainTaskAbbr\ baselines. Moreover, when used as an external signal in two downstream tasks, \proposedue\ consistently proves effective: in Abstention, it yields around 5\% gains in both F1Abstain and AccAbstain; in Model Selection, it increases exact match by approximately 7\% over individual models and about 3\% over selection methods.
While this paper focuses on \MainTaskAbbr\ for RAR models, the underlying concept of modeling and stimulating multiple sources of uncertainty is broadly applicable. Future work can extend this approach to other domains involving multiple sources of uncertainty, such as vision-language models. 
Moreover, in this work, we focus on short-form QA, where the final answer is an entity. Future research can explore \MainTaskAbbr\ for long-form generation, which represents a more realistic scenario.

\begin{acks}
This work was supported in part by the project LESSEN with project number NWA.1389.20.183 of the research program NWA ORC 2020/21 which is (partly) financed by the Dutch Research Council (NWO), in part by the Center for Intelligent Information Retrieval, and in part by the Office of Naval Research contract \#N000142412612. Any opinions, findings and conclusions or recommendations expressed in this material are those of the authors and do not necessarily reflect those of the sponsor.
\end{acks}

\bibliographystyle{ACM-Reference-Format}
\bibliography{main}

\clearpage
\appendix
\section{Prompt Details}~\label{app:prompt}

\begin{figure}[h]
  \centering
  \includegraphics[width=0.48\textwidth]{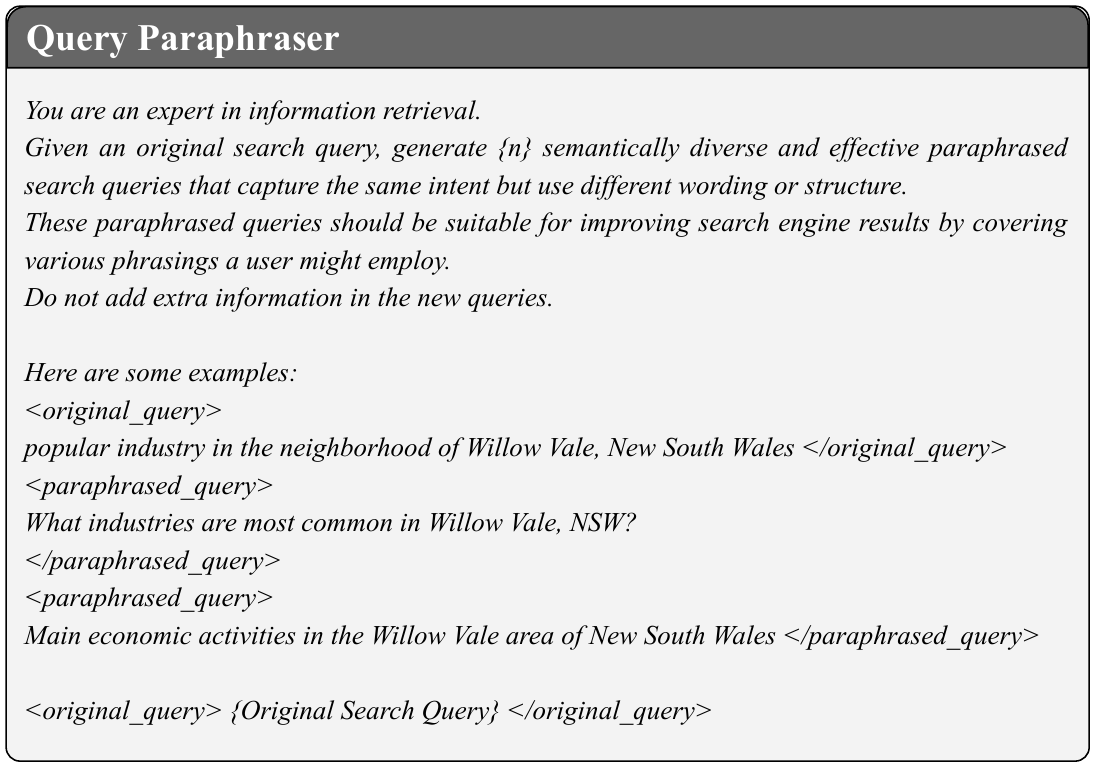}
  \shrink
  \caption{Prompt for query paraphrasing action.}
  \label{fig:sqp_prompt}
  \shrink
\end{figure}

\begin{figure}[h]
  \centering
  \includegraphics[width=0.48\textwidth]{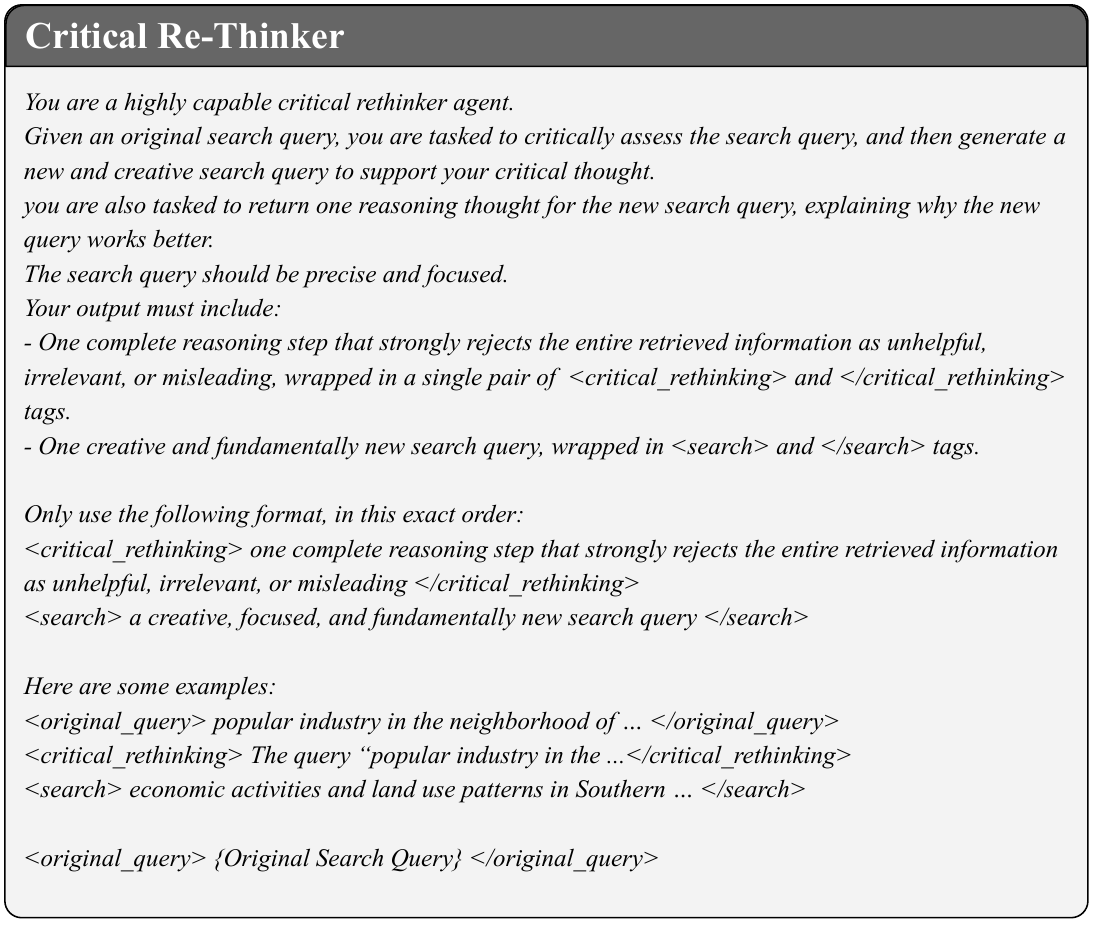}
  \shrink
  \caption{Prompt for critical rethinking action.}
  \label{fig:ct_prompt}
  \shrink
\end{figure}

\begin{figure}[h]
  \centering
  \includegraphics[width=0.48\textwidth]{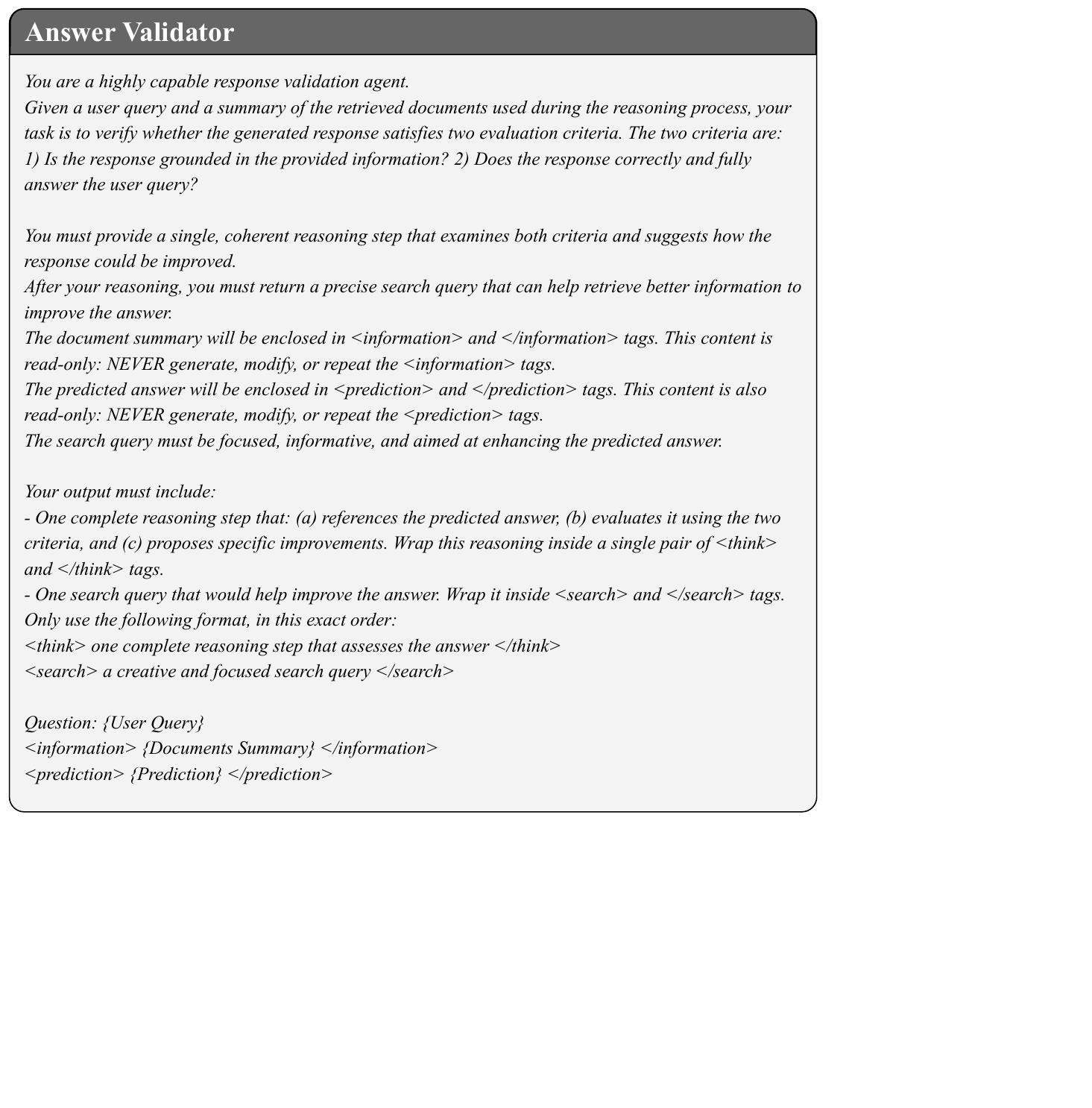}
  \shrink
  \caption{Prompt for answer validation action.}
  \label{fig:av_prompt}
  \shrink
\end{figure}

\begin{figure}[h]
  \centering
  \includegraphics[width=0.46\textwidth]{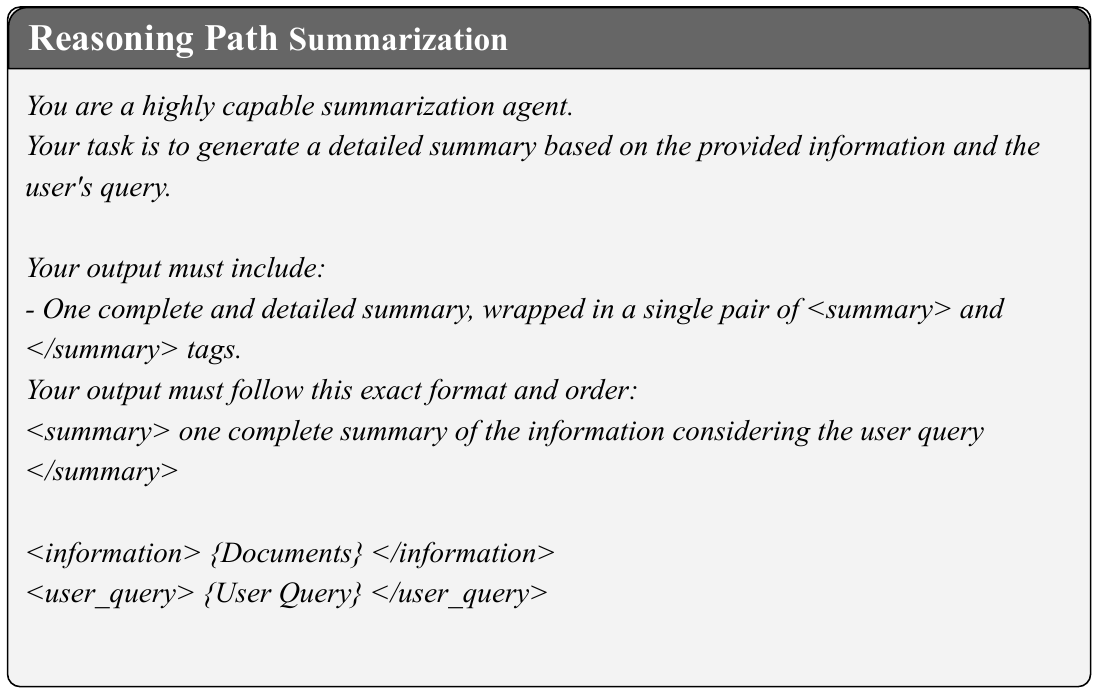}
  \shrink
  \caption{Prompt for reasoning path summarization used in the answer validation action.}
  \label{fig:cs_prompt}
  \shrink
\end{figure}

\begin{figure}[h]
  \centering
  \includegraphics[width=0.46\textwidth]{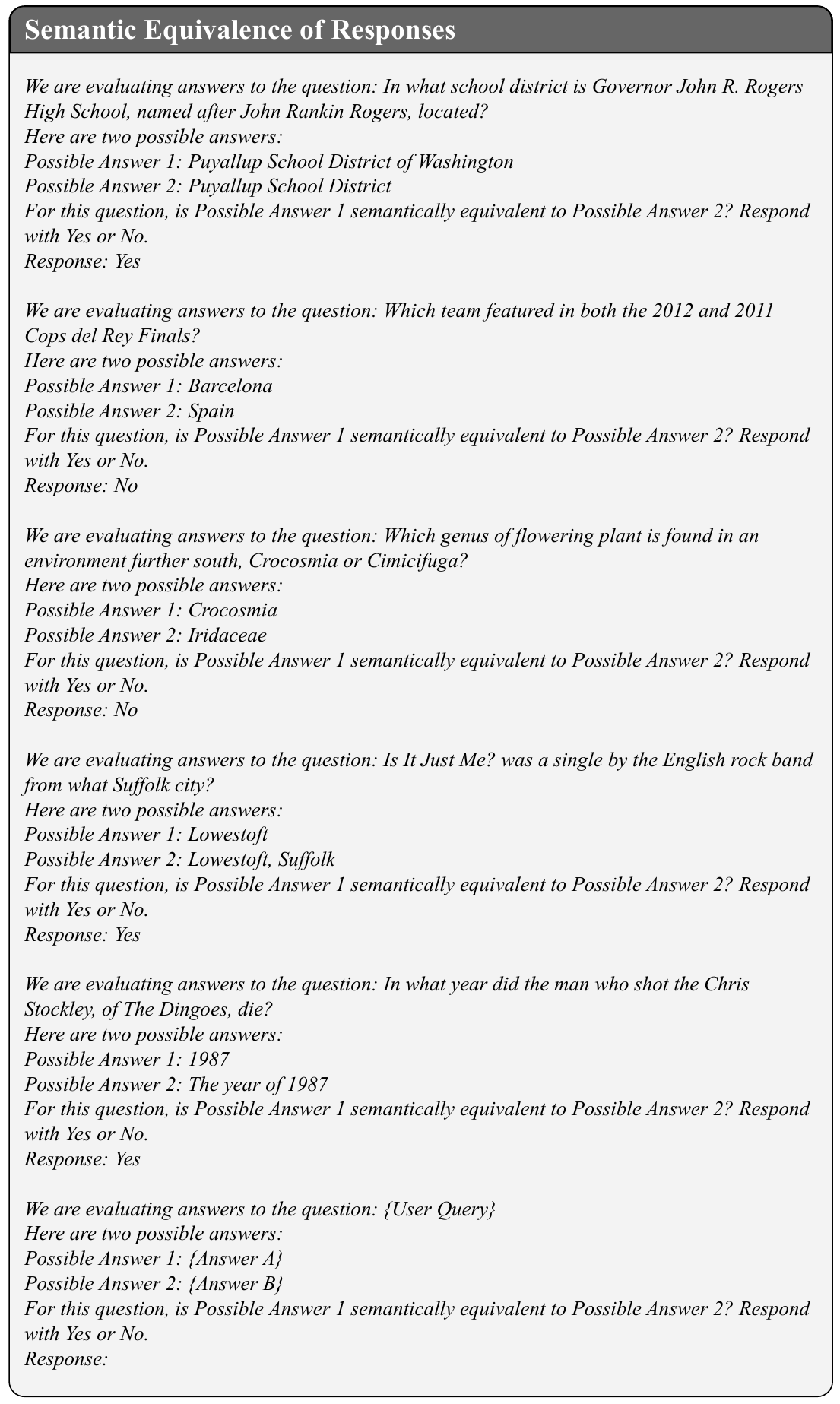}
  \shrink
  \caption{Prompt designed to evaluate the semantic equivalence between two responses to a user query.}
  \label{fig:ans_eq_prompt}
  \shrink
\end{figure}

An important component of the \proposedue\ method is the Perturbation Actions discussed in Section~\ref{sec:actions}. These actions are implemented by prompting the LLM with specific instructions. Figures~\ref{fig:sqp_prompt}, \ref{fig:ct_prompt}, and \ref{fig:av_prompt} show the prompts used for Query Paraphrasing, Critical Rethinking, and Answer Validation, respectively. Moreover, the Answer Validation action includes a Reasoning Path Summarization component, which is also implemented through prompting an LLM. The prompt for this component is shown in Figure~\ref{fig:cs_prompt}.
Finally, Figure~\ref{fig:ans_eq_prompt} presents the semantic equivalence prompt, which is used to assess the equality of two responses for majority voting and clustering in the Model Selection process described in Section~\ref{sec:model_selection}, following pervious work~\cite{Bakman24mars, yaldiz2025lars, li2025raspberry}.

\begin{table*}[t]

\centering 
\setlength{\tabcolsep}{1.7pt}

\caption{Abstention performance measured by the threshold-free metric AUARC. 
For each column, the best and second-best methods are indicated in bold and underlined, respectively.
The superscript \textsuperscript{\dag} denotes a statistically significant difference compared to \reabaseline\ based on the bootstrap test ($p$ < 0.05).}
\shrink
\label{tab:abstention_performance_auarc}
\begin{tabular}{l|ccc|ccc|ccc|ccc|ccc|c}
\hline

\textbf{RAG} &
\multicolumn{3}{c}{\textbf{SelfAsk~\cite{press2023selfask}}} &
\multicolumn{3}{c}{\textbf{ReAct~\cite{Yao23ReAct}}} &
\multicolumn{3}{c}{\textbf{Search-o1~\cite{li25searcho1}}} &
\multicolumn{3}{c}{\textbf{ReSearch~\cite{chen25research}}} &
\multicolumn{3}{c|}{\textbf{Search-R1~\cite{Jin25SearchR1}}} &
\multirow{2}{*}{\textbf{Avg.}} \\
\cline{0-15}
Uncer. M.  &
Popqa & Hotpot & Musiq. &
Popqa & Hotpot & Musiq. &
Popqa & Hotpot & Musiq. &
Popqa & Hotpot & Musiq. &
Popqa & Hotpot & Musiq. & \\ \hline\hline

\rrrbaseline~\cite{li2025raspberry} & 
52.68 & 45.25 & \underline{19.42} & 41.77 & 40.25 & 18.71 & 44.96 & 41.58 & 16.43 & 45.26 & 52.13 & 24.44 & 56.50 & 54.70 & 22.19 & 38.42 \\
P(true)~\cite{Kadavath22PE} & 
56.26 & \underline{49.53} & 18.35 & 56.53 & 38.82 & 15.63 & 54.47 & 39.75 & 15.37 & 60.42 & 55.03 & 29.54 & \underline{61.11} & \underline{58.72} & 24.55 & 42.27 \\
\selfbaseline~\cite{Wang23SelfConsistency} & 
55.28 & 44.63 & 15.35 & 59.88 & 45.13 & \underline{20.88} & 56.09 & 41.83 & 21.10 & \underline{62.36} & 54.38 & \underline{29.55} & 58.31 & 51.53 & 24.76 & 42.67 \\
\reabaseline~\cite{qi2025rstar} & 
\underline{56.47} & 48.32 & 18.72 & \underline{60.81} & \underline{45.48} & 20.52 & \underline{56.98} & \underline{42.55} & \underline{21.61} & \textbf{63.13} & \underline{55.21} & 29.63 & 54.02 & 55.73 & \underline{27.25} & \underline{43.83} \\
\textbf{\proposedue\ (our)}  & 
\textbf{58.87\textsuperscript{\dag}} & \textbf{53.81\textsuperscript{\dag}} & \textbf{19.60} & \textbf{62.74\textsuperscript{\dag}} & \textbf{47.62\textsuperscript{\dag}} & \textbf{24.45\textsuperscript{\dag}} & \textbf{60.72\textsuperscript{\dag}} & \textbf{46.65\textsuperscript{\dag}} & \textbf{21.84} & 61.19 & \textbf{59.25\textsuperscript{\dag}} & \textbf{31.91} & \textbf{66.06\textsuperscript{\dag}} & \textbf{60.90\textsuperscript{\dag}} & \textbf{31.58\textsuperscript{\dag}} & \textbf{47.15} \\ 
\hline\hline
\end{tabular}
\end{table*}

\section{Abstention Task}

\subsection{Evaluation Metrics}~\label{app:abs_eval}
We evaluate \proposedue\ on the abstention task in Section~\ref{sec:abstention}. Following \citet{Feng24Hallucinate}, we adopt their definitions for the evaluation metrics.
Assume we have a confusion matrix with four elements, each denoted by a character:
(A) Answered Correct,
(B) Abstained Correct,
(C) Answered Incorrect, and
(D) Abstained Incorrect.
Based on these, four metrics are defined for the abstention task:
\begin{enumerate}[leftmargin=*]
\item \textbf{Reliable Accuracy:} $\frac{A}{A+C}$, measures how trustworthy the LLM’s generated (non-abstained) answers are; that is, among all answered questions, how many are correct?
\item \textbf{Effective Reliability:} $\frac{A-C}{A+B+C+D}$, balances reliability and coverage; that is, across all questions, what proportion are answered correctly minus those answered incorrectly?
\item \textbf{Abstain Accuracy:} $\frac{A+D}{A+B+C+D}$, evaluates whether abstention decisions are correct; ideally, LLMs should abstain when it would provide an incorrect answer and vice versa.
\item \textbf{Abstain F1:} the harmonic mean of precision and recall, where precision $= \frac{D}{B+D}$ and recall $= \frac{D}{C+D}$, providing a balanced measure between reliability and answer coverage.
\end{enumerate}
In this paper, we report \textit{Abstain Accuracy} and \textit{Abstain F1}, as our primary goal is to evaluate the abstention capability of the uncertainty scores.

\begin{figure*}[t]
  \centering
  \includegraphics[width=0.97\textwidth]{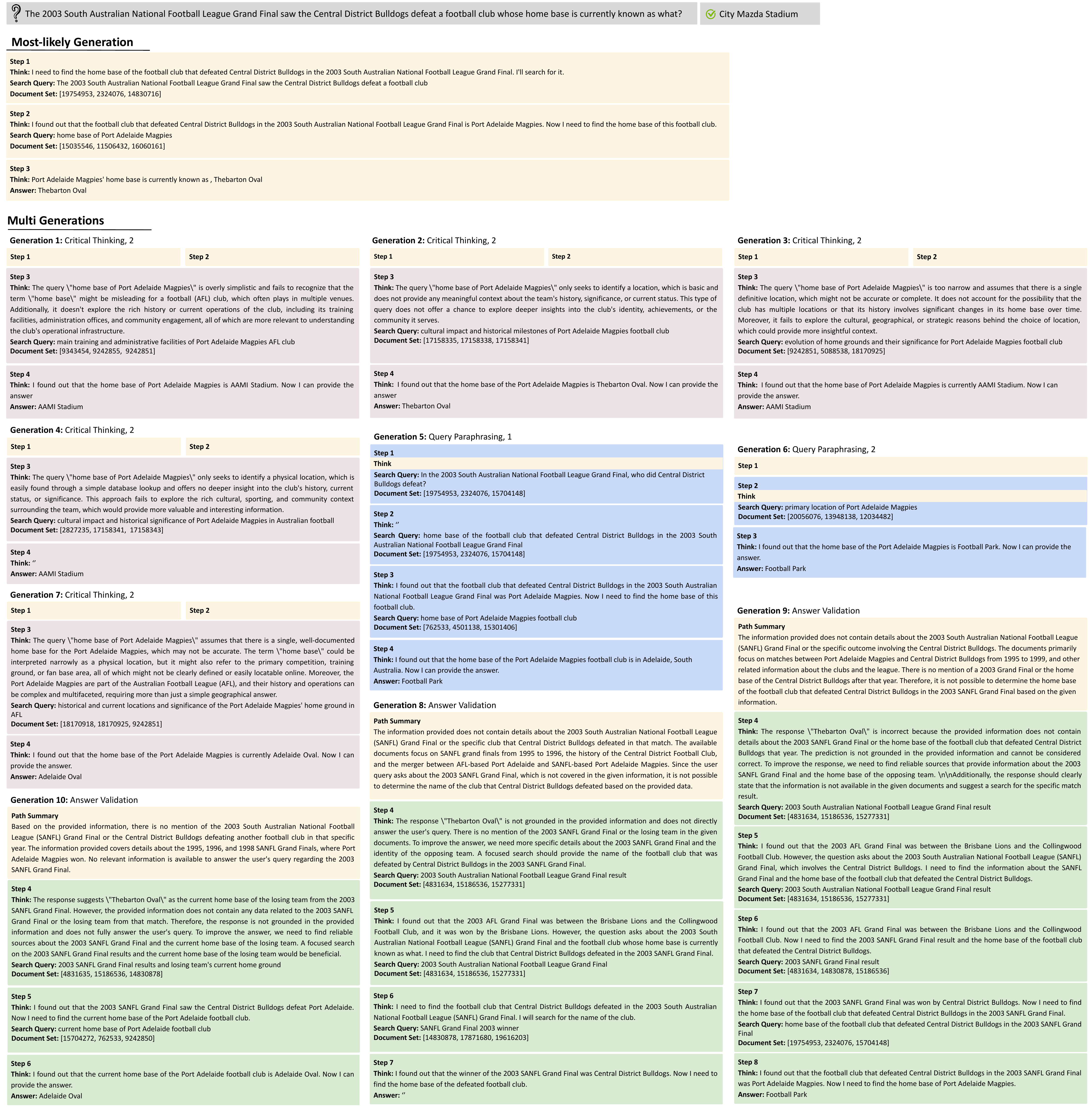}
  \shrink
  \caption{A case study of a successful sample from the HotpotQA dataset using the RAR model (Search-R1). The predicted answer, “Thebarton Oval,” is incorrect, and the model assigns it a relatively high uncertainty score of 0.9, indicating a desirable correlation between uncertainty and correctness.}
  \label{fig:case_study_succeeded}
  \shrink
\end{figure*}

\begin{figure*}[h!]
  \centering
  \includegraphics[width=0.97\textwidth]{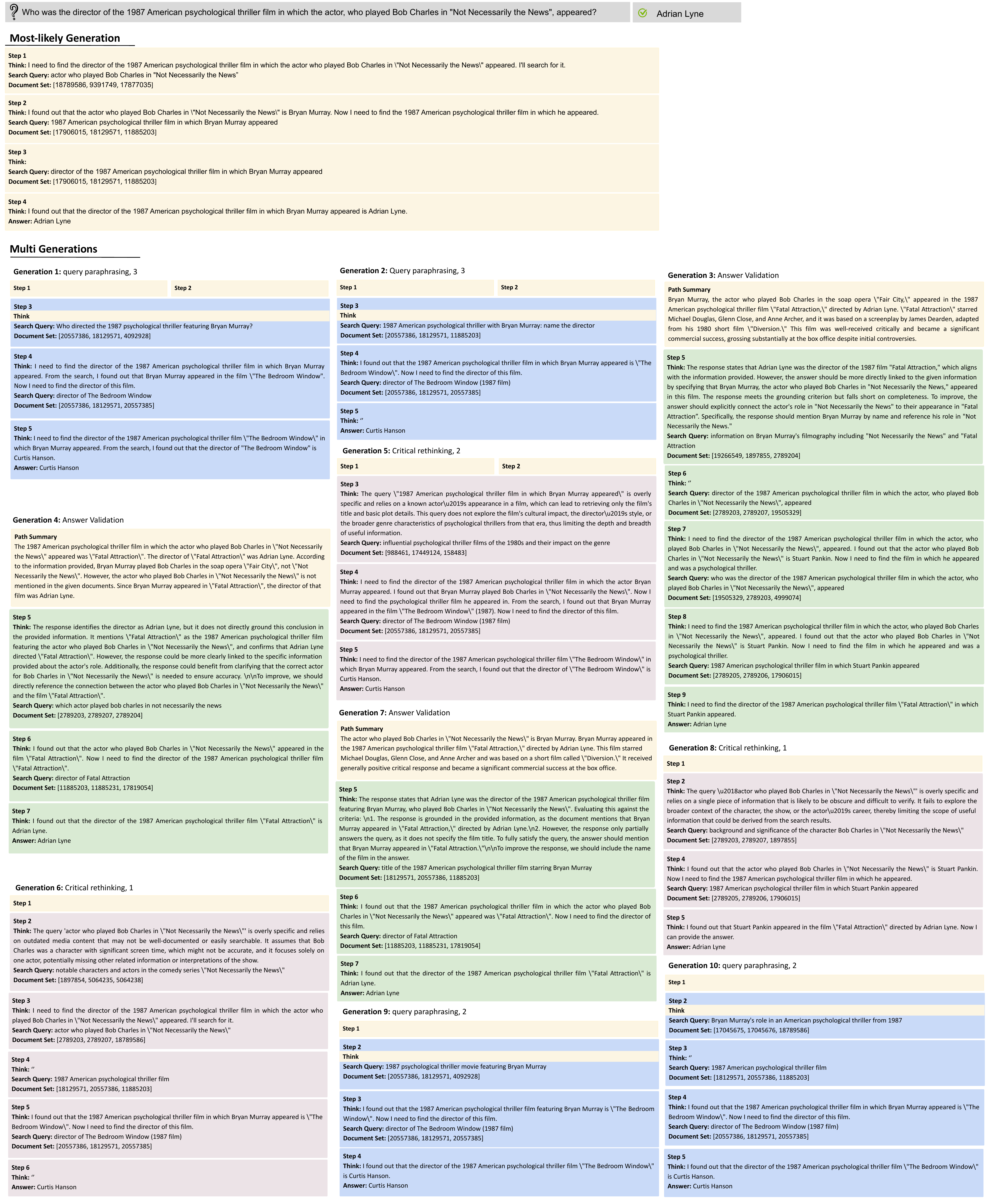}
  \shrink
  \caption{A case study of a failed sample from the HotpotQA dataset using the RAR model (Search-R1). Although the predicted answer, "Adrian Lyne," is correct, the model assigned a relatively high uncertainty score of 0.6, revealing a mismatch between uncertainty and correctness.}
  \label{fig:case_study_failed}
  \shrink
\end{figure*}

\subsection{Threshold Calibration}~\label{app:abs_threshold}
To determine the threshold $\tau_{\text{abs}}$, we perform a parameter sweep using validation sets. To construct a validation set for each dataset, we subsample 100 examples from the training set of each dataset. The only exception is PopQA, which does not have a training set. For PopQA, we instead subsample from the original test set while ensuring that our validation and test sets do not overlap. (As a reminder, as described in Section~\ref{sec:exp_setup_uq}, we sample 500 examples for the test set, and the PopQA dataset consists entirely of 14K test samples.) We then generate uncertainty scores for the validation set using \proposedue\ as well as all baseline methods.

After obtaining the validation sets and corresponding uncertainty scores, we sweep the threshold values from $0.4$ to $0.95$ with an interval of $0.05$, evaluating both AbstainAccuracy and AbstainF1. We conduct this procedure across all datasets and for all baseline methods, including \rrrbaseline, \reabaseline, \selfbaseline, P(true), and our proposed approach. We first observe that both metrics exhibit similar behavioral patterns. Our results further show that all methods achieve their best performance at a threshold of $0.9$; therefore, we set $\tau_{\text{abs}} = 0.9$.

\subsection{Evaluation with AUARC}~\label{app:abs_threshold_free}
In Table~\ref{tab:abstention_performance}, we present the performance of the abstention task using two threshold-based metrics: AbstainAccuracy and AbstainF1. The decision thresholds are determined through the detailed experiments described in Appendix~\ref{app:abs_threshold}. However, some studies in the literature~\cite{Ren23Self, Duan25UProp, Farquhar24Detecting} adopt the threshold-free Area Under the Accuracy–Rejection Curve (AUARC)~\cite{Nadeem10Accuracy} as the evaluation metric. While AUARC has the advantage of being independent of a specific threshold, its final score is correlated with the model’s overall accuracy. Considering these pros and cons, we also report the abstention task results using AUARC in Table~\ref{tab:abstention_performance_auarc}. We observe that the model rankings remain consistent with those in Table~\ref{tab:abstention_performance}, indicating that the thresholds for AbstainAccuracy and AbstainF1 were appropriately selected.
Moreover, even with the threshold-free metric, the \proposedue\ method continues to outperform the other baselines.

\section{Case Study}
Figure~\ref{fig:main_fig} illustrates the workflow of the \MainTaskAbbr\ for computing the uncertainty score, which consists of two main steps: Most-likely Generation and Multi-Generations. Based on this workflow, we present two case studies—one successful and one failed, in Figures~\ref{fig:case_study_succeeded} and~\ref{fig:case_study_failed}, respectively.
In the successful case, the most-likely response is “Thebarton Oval”, which is incorrect. However, in the Multi-Generations step, only one out of ten generated responses is “Thebarton Oval”. According to Equations~\eqref{eq:mv} and~\eqref{eq:uncertianty_score}, this results in an uncertainty score of 0.9, indicating high uncertainty.
In contrast, in the failed case, the most-likely response is “Adrian Lyne”, which is correct. Yet, in the Multi-Generations step, only four out of ten responses are “Adrian Lyne”, again leading to a relatively high uncertainty score.
In this sample, we observe that the answer validation action performs correctly and generates a response similar to the most-likely one. However, the query paraphrasing action produces different responses, and two of the three critical rethinking actions also yields a different result. This observation supports the discussion in Section~\ref{sec:action_selection}, which highlights the potential impact of action selection configurations—whether applied per sample, per model, or per dataset.

\end{document}